\documentclass[pra, twocolumn, superscriptaddress, amsmath, amssymb, floatfix]{revtex4}

\usepackage[utf8]{inputenc}
\usepackage[english]{babel}
\usepackage{graphicx}
\usepackage{braket}
\usepackage{wrapfig}
\usepackage{hyperref}
\usepackage{bm}
\usepackage{subfigure}
\usepackage{color}
\usepackage{soul}

\def\rme{{\rm {e}}}
\def\rmi{{\rm {i}}}
\def\tr{{\rm{Tr}}}

\def\be{\begin{equation}}

\def\ee{\end{equation}}
\def\ba{\begin{eqnarray}}
\def\ea{\end{eqnarray}}

\newcommand{\lio}{\mathcal{L}}

\newcommand{\h}{\mathcal{H}}
\newcommand{\E}{\mathcal{E}}
\newcommand{\ada}{\hat{a}^{\dagger}}
\newcommand{\aaa}{\hat{a}}
\newcommand{\bda}{\hat{b}^{\dagger}}
\newcommand{\bbb}{\hat{b}}
\newcommand{\sgn}{\mathop{\mathrm{sgn}}}

\begin{document}

\title{Photon transport in a dissipative chain of nonlinear cavities}
\author{Alberto Biella}
\affiliation{NEST, Scuola Normale Superiore \& Istituto Nanoscienze-CNR, I-56126 Pisa, Italy}

\author{Leonardo Mazza}
\affiliation{NEST, Scuola Normale Superiore \& Istituto Nanoscienze-CNR, I-56126 Pisa, Italy}

\author{Iacopo Carusotto}
\affiliation{INO-CNR BEC Center \& Dipartimento di Fisica, Universit\`a di Trento, I-38123 Povo, Italy}  

\author{Davide Rossini}
\affiliation{NEST, Scuola Normale Superiore \& Istituto Nanoscienze-CNR, I-56126 Pisa, Italy}

\author{Rosario Fazio}
\affiliation{NEST, Scuola Normale Superiore \& Istituto Nanoscienze-CNR, I-56126 Pisa, Italy}

\date{\today}

\begin{abstract}
By means of numerical simulations and the input-output formalism we study photon transport through a chain of coupled 
nonlinear optical cavities  subject to uniform dissipation. Photons are injected from one end of the chain by means of a 
coherent source. The propagation through the array of cavities is sensitive to the interplay between the photon hopping 
strength and the local non-linearity in each cavity. We characterize photon transport by studying the populations and the 
photon correlations as a function of the cavity position. When complemented with input-output theory, these quantities provide 
direct information about photon transmission through the system. The position of single- and multi-photon resonances directly 
reflects the structure of the many-body energy levels. This shows how a study of transport along a coupled cavity array can 
provide rich information about the strongly correlated (many-body) states of light even in presence of dissipation.
The numerical algorithm we use, based on the time-evolving block decimation scheme adapted to mixed states, allows us to 
simulate large arrays (up to sixty cavities). The scaling of photon transmission with the number of cavities does depend on the 
structure of the many-body photon states inside the array.
\end{abstract}

\maketitle

\section{Introduction}

In the recent years, Coupled Cavity Arrays (CCAs)~\cite{hartmann2006, greentree2006, angelakis2007}, have been 
put forward as a very suitable playground for the investigation of quantum many-body phenomena in photonics systems. 
Due to their flexibility in the design, the possibility to control their dynamics (through the choice of the couplings and external 
drive) and the local accessibility of individual cavities, these systems have been proposed as possible implementations of 
a quantum simulator. A rather comprehensive account of the large body of work in this field can be found in 
Refs.~\cite{hartmann2008, tomadin2010, houck2012, carusotto2013}.  The experimental requirements are quite challenging, 
however in the last two years there have been very interesting progresses~\cite{underwood2013, Abbarchi13, Toyoda13}.

An important ingredient determining the dynamics of a cavity array is the competition between photon hopping and the 
non-linearity present in each cavity, due to the coupling to a few-level system as for example in the Jaynes-Cummings model. 
Whereas tunnelling between neighbouring cavities tends to delocalise the photons, the presence of the non-linearity, 
on the contrary suppresses number fluctuations, thus opposing to delocalisation. In the (hypothetical) absence of photon 
losses  this competition would lead to a (thermo)dynamics similar to that of the Bose-Hubbard model. The properties of 
cavity arrays in this regime have been carefully scrutinised in the recent literature, see e.g. the reviews~\cite{hartmann2008, 
tomadin2010, houck2012}. The phase diagram in the one-dimensional case, related to the present study, has been 
determined by means of density matrix renormalization group in~\cite{rossini2007}. 

\begin{figure*}
\includegraphics[width=0.70\textwidth]{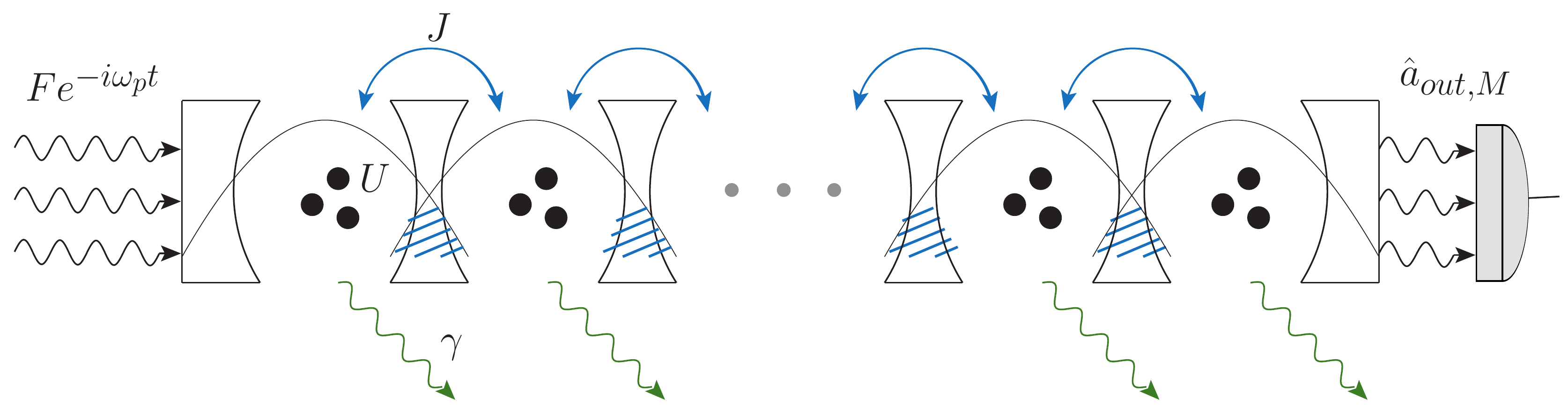}
\caption{A sketch of the one-dimensional cavity array.  Neighboring cavities are coupled by photon hopping. Nonlinearities in the cavities 
	may produce an effective repulsion between the photons leading to an anharmonic spectrum. We consider a Kerr-like nonlinearity. 
	Photons in the cavities have a finite lifetime therefore the cavities are pumped with an external coherent drive. Here we suppose that 
	only the leftmost cavity is pumped, in order to study photon transport 
    	through the system.}
  \label{fig:cavities}
\end{figure*}

The presence of the unavoidable photon leakage would make the long-time dynamics completely trivial in the absence of 
an external drive that refills photons into the cavities.  This additional competition, between losses and  external pumping, 
makes the dynamical behaviour of these systems particularly rich. The interplay of (coherent/incoherent) driving and (incoherent) 
photon losses can be both observed in the transient and in the steady-state (long-time) regime.  In this work we will be interested in 
the Non-Equilibrium Steady State (NESS). 

Only very recently the many-body non-equilibrium dynamics of cavity arrays started to be addressed --see, {\it e.g.},  
Refs.~\cite{carusotto2009, tomadin2010b, Hartmann10, nissen2012, grujic2012, grujic2013, jin2013, yuge2013, RuizRivas14}-- 
and several different properties both of the transient and steady-state regimes were highlighted. These include the spectroscopy of many-body photon states~\cite{carusotto2009, nissen2012, grujic2012, grujic2013, RuizRivas14},  crystallisation of 
photons~\cite{Hartmann10}, instabilities induced by pulsed driving~\cite{tomadin2010b}, steady state critical phenomena~\cite{jin2013, 
yuge2013}. There is by now a compelling evidence that cavity arrays are naturally designed to become open-system quantum simulators.  

Very interesting many-body effects should emerge in photon transport as well.  Most of the attention so far has been devoted to 
the transmission of one or two photons. In this regime, transnport in one-dimensional 
systems has been already studied in a variety of situations, in the presence of  a single two-level system (see, for example, 
Refs.~\cite{shen2005-2009, chang2007, longo2010, felicetti2014}), with extended nonlinearities~\cite{hafezi2011} and in linear cavity 
arrays when the frequency of one or two cavities is tunable~\cite{liao2010}.  

Moving away from the regime of few-photon transmission, it is natural to expect that  the formation of many-body states of photons 
will considerably affect transport as well; to the best of our knowledge, this effect has not been investigated so far. It is important to stress from 
the beginning that it is not obvious how strong correlations play a role in photon transport. For a single cavity, the most striking effect is 
the photon blockade~\cite{imamoglu97,rebic99,kim99,birnbaum05}.  The presence of a single photon in a cavity, driven by an external coherent 
source,  will block the passage of a second photon because of the strong non-linearities present in the cavity itself. 
How is the photon blockade modified in an extended system? This is one of the questions that we will address.

We make a 
first step in this direction by addressing the problem of photon transport through a chain of nonlinear cavities in presence of dissipation. 
The setup we consider is sketched in Fig.~\ref{fig:cavities}. One end of the array is driven by a coherent laser source. We are interested 
in studying the light emerging at the other end of the array, linking its properties to the presence of complex many-body states 
in the array itself.  Our interest starts from Ref.~\cite{carusotto2009}, where it was shown that the steady state of a ring of nonlinear 
cavities, uniformly pumped, is reminiscent of a strongly correlated Tonks-Girardeau (TG) gas of impenetrable bosons.
Inspired by this work, here we explore the impact of  strongly correlated many-body states on transport.

Furthermore, it is worth mentioning at this stage that this system is also relevant for the understanding of single photon sources based on 
passive photonics devices~\cite{liew2010, ferretti2013}.  Changing a bit the perspective, the configuration we propose, 
can be view as an alternative scheme to detect and quantify the presence  of strongly correlated states of light based on transport. 
Usually, this kind of spectroscopic analysis is performed driving the whole array and studying the near-field and the far-field pattern.

In this work we mainly focus in the regime where on-site nonlinearities are much stronger than photon hopping. First we concentrate 
in the limit of impenetrable bosons: in this regime, repulsively interacting bosons form a Tonks Girardeau (TG) gas 
and behave as non-interacting fermions~\cite{girardeau1960}. Such limit is inaccessible with state-of-the-art experiments, 
but it is interesting in view of the rarity of exact solutions in many-body problems,  and serves as a helpful benchmark for approximation 
methods in many-body theory. 

Determining the steady state, {\it i.e.}, the long-time limit of a many-body Lindblad equation, is a formidable task. In many-body open 
quantum systems the unavoidable exponential growth of the Hilbert space with the system size merge with the need to represent 
mixed states, leading to a huge number of degrees of freedom to be taken into account. By means of an extension 
of the time-dependent density-matrix renormalization group  to open systems~\cite{zwolak2004,verstraete2004}, we are able to attack 
this problem and study a large number (up to sixty) of  coupled cavities. Our numerical simulations are validated by means of an 
analytic approach based on effective models which take into account only few relevant degrees of freedom.  The combination of 
(essentially) exact numerical methods together with the judicious construction of effective models allows to considerably enrich the 
understanding of the underlying physics.

The paper is organized as follows. In the next section we discuss in details the model for the driven/dissipative coupled cavity chain 
of Fig.~\ref{fig:cavities}. We will also discuss the basics of the input-output formalism that will allow to compute transport properties.
Sections~\ref{sec:results} and~\ref{sec:finite} are devoted to the presentation of our results: in Sec.~\ref{sec:results}, we concentrate 
in the case of hard core bosons, while the case of finite interaction strength is discussed in Sec.~\ref{sec:finite}. 
Finally, in Sec.~\ref{sec:conclusions} we summarise our conclusions.

\section{The model}
\label{sec:model}

We study transport properties of a one-dimensional array of $M$ optical cavities, coupled by photon tunnelling, 
each one displaying an optical nonlinearity of the Kerr type.
After tracing out the environmental degrees of freedom, the time evolution of the system 
density matrix is ruled by a master equation in the Lindblad form 
\be
\dot{\rho}=-\frac\rmi\hbar[\hat \h, \rho] + \lio[\rho] ,
\label{lindblad}
\ee 
where the first term describes the coherent unitary time evolution, and the Lindblad term takes into account the damping. 
In the rest of the paper we will consider independent photon losses from each cavity as the only dissipation channel.

Assuming that the spacing between the modes of each cavity is larger than any other 
involved energy scale, we can write the system Hamiltonian considering only one mode for each cavity
\be
\hat \h_0 = \hbar\omega_0\sum_{i=1}^M \ada_i \aaa_i + \hbar U\sum_{i=1}^M\ada_i \ada_i \aaa_i \aaa_i 
-\hbar J\sum_{i=1}^{M-1}(\aaa_{i}\ada_{i+1}+\text{h.c.}),
\ee 
where $\aaa_i$ ($\ada_i$) are bosonic photon annihilation (creation) operators associated with 
the $i$-th cavity of the chain with resonance frequency $\omega_0$ which obey the canonical commutation 
relations ($[\aaa_i,\ada_j]=\delta_{i,j}$, $[\aaa_i,\aaa_j]=0$), $J$ is the hopping rate and $U$ sets the scale of the Kerr nonlinearity.

The system is coherently driven by an incident (monochromatic) laser beam. In the setup we are interested in only the first cavity 
is coherently driven. In the input-output formalism~\cite{milburn}  the equation of motion of the field operator 
in the first cavity (in the Heisenberg picture) is modified as follows
\be
\label{inout}
\partial_t \aaa_1 (t)= \frac\rmi\hbar [\hat\h_0, \aaa_1(t)] -\frac\kappa2 \aaa_1(t) + \sqrt\kappa \ \aaa_{in,1}(t),
\ee
where $\aaa_{in,1}$ is the (laser) input field shined on the first cavity and $\kappa$ is the coupling between the cavity 
mode $\aaa_1$ and the laser field. The laser is in a coherent state and then can be written as the input relative to 
the vacuum $\aaa_{in,1}^{vac}$ displaced by $\alpha$ and rotating at the pump frequency $\omega_p$ ($\aaa_{in,1} = 
\aaa_{in,1}^{vac} + \alpha \ \rme^{-\rmi\omega_p t - \rmi\phi}$). By substituing it in Eq.~\eqref{inout} we get
\begin{eqnarray}
\label{inout2}
\partial_t \aaa_1(t) &=& \frac\rmi\hbar [\hat\h_0, \aaa_1(t)] -\frac\kappa2 \aaa_1(t) \nonumber \\
                              &+& \sqrt\kappa (\ \aaa_{in,1}^{vac}(t) + \alpha \ \rme^{-\rmi\omega_pt - \rmi\phi}),
\end{eqnarray}
The last term on the r.h.s. of Eq.~\eqref{inout2} can be taken into account at the Hamiltonian level 
by adding a term to $\hat \h_0$
\be
\hat \h = \hat \h_0 + \hbar \ [F(t)\ada_1 + F^*(t)\aaa_1],
\ee
where $F(t)=F \rme^{-\rmi\omega_p t }$ with $F=|F(t)|=\sqrt\kappa\alpha$ and $\phi=\pi/2$.
After transforming to a frame rotating with the laser frequency $\omega_p$, 
the Hamiltonian takes the form 
\be
\hat \h = \hat \h_0 -\hbar\omega_p \sum_{i=1}^M \ada_i \aaa_i + \hbar  F \ (\ada_1 + \aaa_1) .
\ee
The second and third terms in Eq.(\ref{inout2}) take into account the noise induced from this input-output channel and can 
be safely neglected because the main noise source is due to the uniform photon loss at rate $\gamma$, that is uniform for all the cavities ($\kappa/\gamma\ll1$). 
The corresponding Lindblad term reads
\be
\label{lindbladianterm}
\lio[\rho] =\frac\gamma2\sum_{i=1}^M(2\aaa_i\rho\ada_i-\ada_i\aaa_i\rho-\rho\ada_i\aaa_i).
\ee
The non-trivial competition between unitary time evolution and Lindblad dissipation 
leads to NESS given by the stationary point of the above master equation ($\dot{\rho}=0$).
Specifically we analyze the population and the statistics of the light transmitted by the array $\aaa_{out,M}$.
Again employing input-output theory one can relate the behavior of $\aaa_{out,M}$ to the field in the last cavity of the array $\aaa_M$.
For example, using the relation $\aaa_{out,M}=\aaa_{in,M}+\sqrt{\kappa'}\aaa_M$ and exploiting the fact that $\aaa_{in,M}$ is just the vacuum, for the population one gets 
\be
\label{inout3}
\braket{\ada_{out,M}\aaa_{out,M}} = \kappa' \ \braket{\ada_M\aaa_M},
\ee
where $k'$ takes into account the coupling of the last cavity to the outside.
The notation $\braket{\hat{O}}$ indicates the expectation value of the 
operator $\hat{O}$ taken in the standard way $\braket{\hat{O}}={\cal Z}^{-1} \ \tr[\rho^{\rm \scriptscriptstyle NESS} \, \hat O]$, 
where ${\cal Z} = \tr \, \rho^{\rm \scriptscriptstyle NESS}$ is the partition function and $\rho^{\rm \scriptscriptstyle NESS}$ 
is the NESS density matrix.
Remarkably, the photon statistic will also be exactly the same as the cavity field~\cite{milburn}.
For this reason in this work we show results about the photon density and the correlations of the field in the $M$-th cavity.
As we did previously for the first cavity, we neglect the noise contribution coming from this input-output channel. However, it would be not a problem to rigorously include such noise terms in our model by modifying the loss rate of the first and $M$-th cavity.
In the following we will fix $\hbar=1$ and work in units of $\gamma$.

As mentioned before, from a computational point of view, the simulation of Eq.~\eqref{lindblad} brings together both 
the complexity due to the exponential growth of the Hilbert space 
with the system size and the mixed states dynamics generated by the non-unitary time evolution. 
Since we are interested in describing large arrays, in order to overcome this issue we 
exploit an algorithm based on the time-evolving block decimation (TEBD) scheme~\cite{vidal2003, vidal2004} 
extended to open systems~\cite{zwolak2004, verstraete2004}. 
This relies on the representation of the density matrix as a matrix product 
operator (MPO) and can be viewed as a generalization of a matrix product state (MPS) for non-pure states. 
In the present work we simulate chains with a number of cavities up to $M=60$. 
The bond-link dimension used is $\chi=100$. 
In our system, for typical values of parameter, this representation allows to capture most 
of the entanglement in the NESS.
In App.~\ref{app:mpo} we recall the basic features of the algorithm in order to give 
immediate meaning to the quantities introduced to obtain accurate numerical simulations. 

In order to gain further insight we will supply the MPO simulation with some 
effective models which are able to capture the main features of the NESS. 
The structure of such effective models is detailed in App.~\ref{app:effmod}.

\section{Transport in the Tonks-Girardeau limit} 
\label{sec:results}
 
\begin{figure}[!t]
  \includegraphics[width=0.45\textwidth]{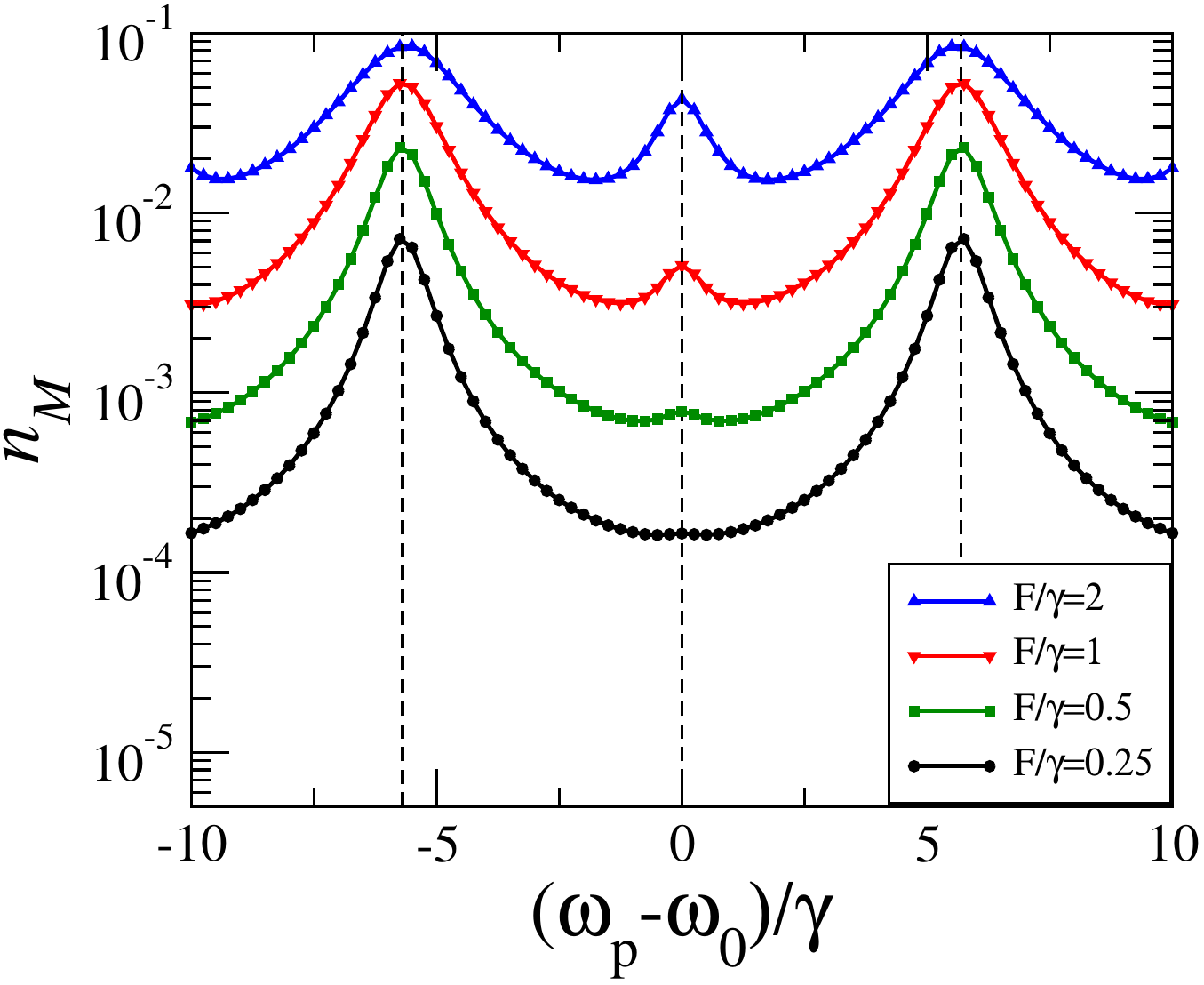}
  \caption{The population in the $M$-th cavity $n_M$ in the NESS as a function of the detuning 
    $(\omega_p-\omega_0)/\gamma$ for different values of the driving strength. 
    The dashed vertical lines are the spectral positions of the peaks in the fermionized 
    limit ($\ket{k_5}$, $\ket{k_5, k_6}$ and $\ket{k_6}$ from left to right). 
    The parameters are $U/J=\infty$, $J/\gamma=20$, $\omega_0/\gamma=1 $ and $M=10$.}
  \label{fig:spectraleft2}
\end{figure}

Let us first consider the limit of impenetrable bosons ($U/J=+\infty$):
in this regime, repulsively interacting bosons form a TG gas and behave 
as non-interacting fermions~\cite{girardeau1960}. 
In Fig.~\ref{fig:spectraleft2}, the population in the $M$-th cavity $n_M=\braket{\ada_M\aaa_M}$ is shown as a 
function of the pump frequency $\omega_p$.  Looking at Fig.~\ref{fig:spectraleft2}, in the detuning range shown, we note the presence of 
two main peaks (symmetrically displaced w.r.t. the zero detuning point) for all the values of the driving strength probed 
and a peak at zero detuning which emerges as the driving strength is increased. In order to understand the nature of this 
peaks it is necessary to analyze the many-body spectrum of $\hat \h_0$.

As shown by Girardeau in Ref.~\cite{girardeau1960}, 
the generic $N$-particle eigenstate of the closed system $\hat \h_0$ can be exactly mapped 
into a fermionic one.
For $N$ bosons, the wave function in real space representation is given by
\be
\label{sim}
\Psi_B (i_1, \dots, i_N) = \prod_{k<j}^N \sgn(i_k-i_j) \ \Psi_F (i_1, \dots, i_N).
\ee
The term $\prod_{k<j}^N \sgn(i_k-i_j)$ ensures that $\Psi_B (i_1, \dots, i_N)$ 
is symmetric under the exchange of any two particles.
In this limit, the eigenstates of $\hat \h_0$ can be simply labeled by the occupation number 
of the single particle eigenstates of the non-interacting Hamiltonian, 
with the prescription that we cannot put more than one particle in each orbital.
The notation $\ket{k^{(1)}, \dots, k^{(N)}}$ indicates the $N$-boson wave function, 
obtained by the symmetrization of the fermionic one via Eq.~\eqref{sim}, 
with one particle in each $k^{(1)}, \dots, k^{(N)}$ orbitals. 
The energy of the $N$-boson wave function $\ket{k^{(1)}, \dots, k^{(N)}}$ is identical 
to the corresponding fermionic one, 
{\it i.e.}, $E=\sum_{\alpha=1}^N \mathcal{E}(k^{(\alpha)})$ where $\E(k^{(\alpha)})= \omega_0 - 2J \cos k^{(\alpha)}$ and the momenta 
$k^{(\alpha=1,\dots,N)}$ are to be chosen in the set $k_n=n\pi/(M+1)$ 
with $n=1,\dots,M$, as imposed by the open boundary conditions.

The peaks in the transmission spectrum are due to the fact that the laser frequency is resonant with some eigenstates of $\hat \h_0$.
Transport, in this regime, occurs through extended many-body photon states of the global system. This is the many-body extension of the 
classic photon-blockade effect in the single driven cavity.  The extended many-body states govern the transport in all the cases we considered 
up to the largest chains of about sixty cavities. It is not obvious that this should be the case since incoherent photon leakage occurs in each cavity 
while driving is only through the first cavity only. Here, and in the rest of the paper, we demonstrate that photon transport can be 
dominated by extended many-body effects. 

Fig.~\ref{fig:spectraleft2} shows that, when the driving strength 
is weak ($F/\gamma\ll1$), only the one-photon states (in the range shown, $\ket{k_5}$ and $\ket{k_6}$) are excited by the pump.
Remarkably, in this driving scheme, all the one-particle states are coupled to the vacuum 
with a matrix element $F_{k_n}=\braket{k_n|\hat \h|0}=F \, \sqrt{2/(M+1)} \, \sin k_n$.
Increasing the driving strength, many-body states start to be excited, 
due to the sequential absorption of $N$ photons from the drive.
This means that a peak in the spectrum relative to the many-body state $\ket{k^{(1)}, \dots, k^{(N)}}$ 
will appear at $\omega_p = \sum_{\alpha=1}^N \mathcal{E}(k^{(\alpha)})/N$,
Specifically, in Fig~\ref{fig:spectraleft2}, the two-photon resonance relative to the state $\ket{k_5, k_6}$ appears at 
$\omega_p=(\mathcal{E}(k_5)+\mathcal{E}(k_6))/2$ as the strength of the pump is increased.
As it is typical in a driven/dissipative scenario, driving and losses imply transitions 
between eigenstates of $\hat\h_0$ with different number of particles and are responsible of the finite linewidth of the resonances. 
Such broadening increases as the driving strength is increased and as a result, the background due to the off resonant excitation of 
the eigenstates of $\hat\h_0$ becomes more  and more important.

\subsection{One-photon resonances}

\begin{figure}[!t]
  \includegraphics[width=0.4\textwidth]{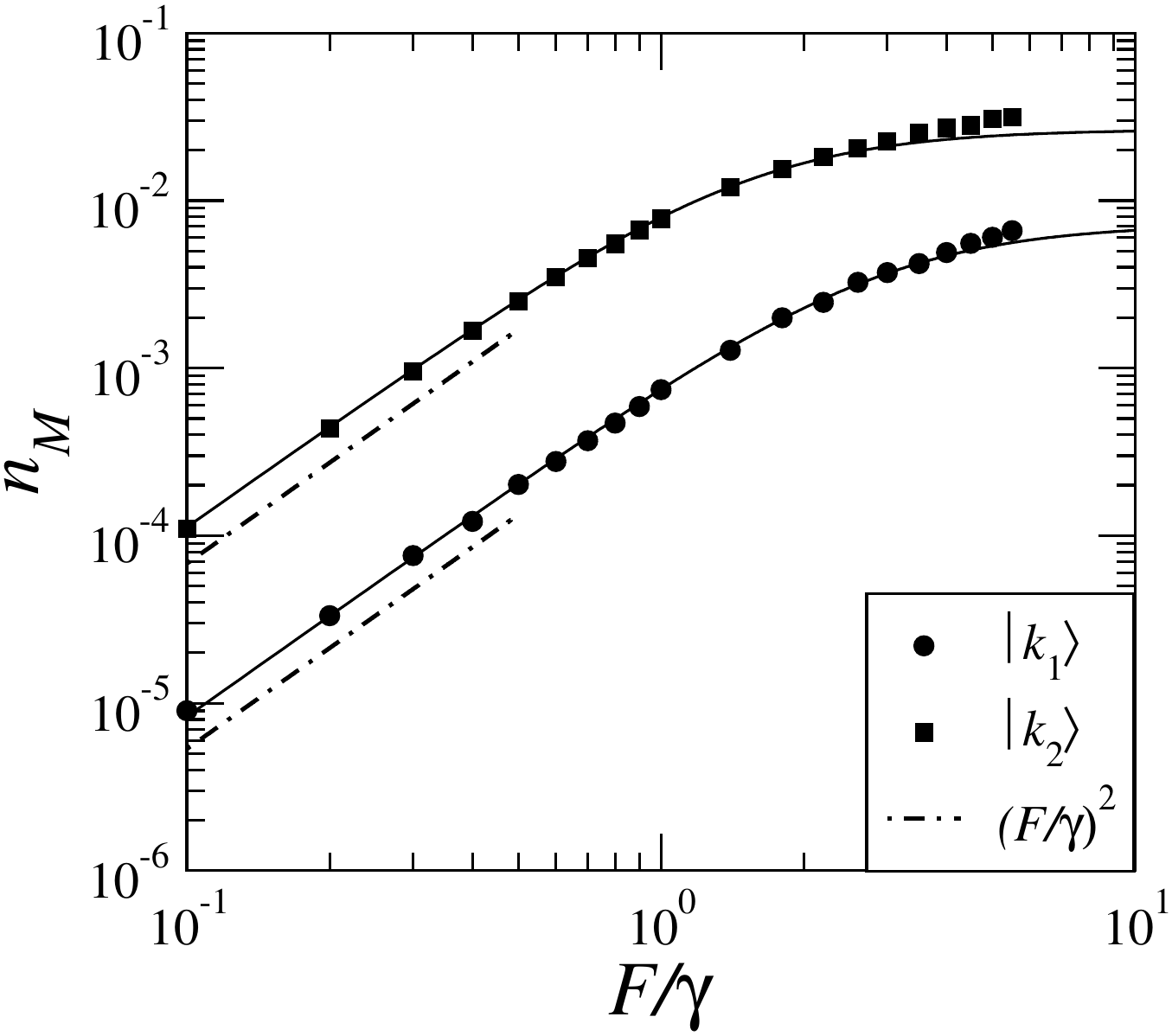}
  \caption{The population in the $M$-th cavity $n_M$ in the NESS as a function 
    of the pump amplitude $F/\gamma$, when the one-photon resonance condition 
    is satisfied for the states $\ket{k_1}$ and $\ket{k_2}$. 
    Symbols denote the numerical data, while solid lines are the outcomes 
    of the OBM (or equivalently of the TLM). 
    Dashed lines indicate a behavior $n_T \propto (F/\gamma)^2$, and are plotted to guide the eye. 
    The parameters are the same as in Fig.~\ref{fig:spectraleft2}.}
  \label{fig:pop1ph2}
\end{figure}

Starting from these initial observations, we want to investigate the structure of the NESS when the driving laser is resonant with a one-photon state 
\be
\ket{k_n}=\sqrt{\frac2{M+1}}\sum_{i=1}^M\sin(k_n i)\ada_i\ket{0}.
\ee
In Fig.~\ref{fig:pop1ph2} the typical behavior of the  population in the $M$-th cavity as a function of the driving strength is shown.

We compare the numerical data (symbols) with the outcome of a truncated 
effective model (solid lines) which involves only the one-body states $\ket{k_m}$ and the vacuum $\ket{0}$.
In this way we take into account both the resonant 
  state $\ket{k_n}$ and the remaining off-resonant one-body states.
In what follows we will refer to this model as the \emph{one-body model} (OBM).
The numerics perfectly agrees with the data of the OBM for $F/\gamma\ll1$, 
where $n_M\propto (F/\gamma)^2$. 
For $F/\gamma\gg1$ the OBM underestimates the population, 
because the off-resonant excitation of many-body states starts to be relevant. 

If the one-photon resonances are well separated in energy with respect to their width, one can further simplify the model considering the resonant state $\ket{k_n}$ and the vacuum only $\ket{0}$, so that an effective two-level model is obtained 
and can be analytically solved (see App.~\ref{app:effmod}). 
We will refer to this model as the \emph{two-level model} (TLM).
This is the case, for example, of the data in Fig.~\ref{fig:pop1ph2},
where the population in the $M$-th cavity in the NESS is shown as a function of the driving strength for an array of ten cavities. In this case  the discrepancy 
between the predictions of the OBM and the TLM is not appreciable.

\begin{figure}[!t]
  \centering
  \includegraphics[width=0.48\textwidth]{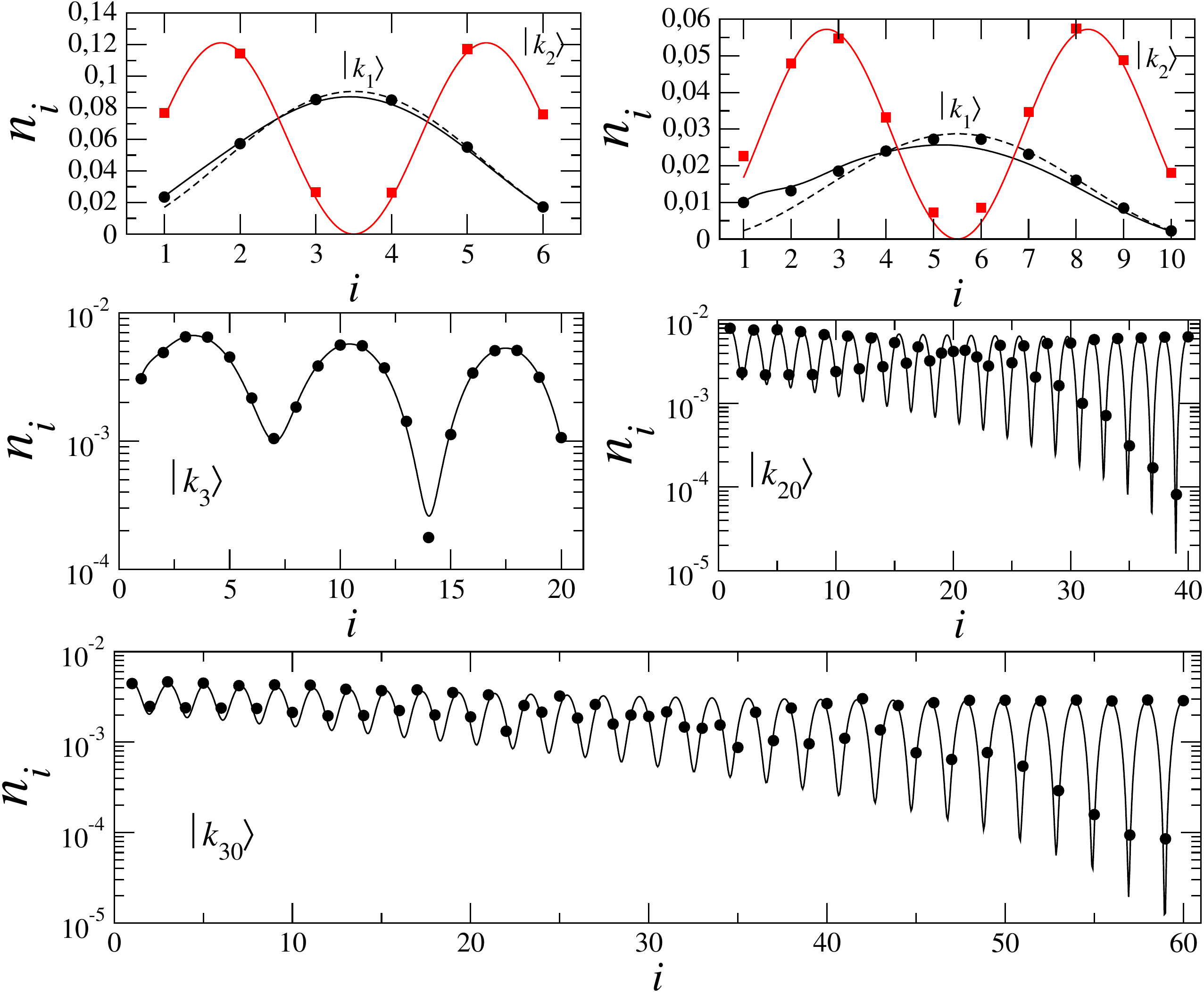}
  \caption{Photon occupations of each site in the NESS, when the one-photon 
    resonance condition is satisfied for different states (as indicated in each panel) and for different system size (respectively $M=6,10,20,40,60$). 
    Symbols denote the numerical data, solid lines are the outcomes 
    of the OBM and dashed lines are the results of the TLM (shown for $M=6,10$). 
    The parameters are the same as in Fig.~\ref{fig:spectraleft2}, but with $F/\gamma=2$ for $M=6,10$ and $F/\gamma=1$ in the other panels.}
  \label{fig:dens1phlarge3}
\end{figure}

In Fig.~\ref{fig:dens1phlarge3} the typical density profile on resonance 
with a one-photon state $\ket{k_n}$ is shown. 
As explained above, for small system sizes  (top panels) the one-photon resonances are well separated in energy compared to their width
and then the contribution to the NESS of the off-resonant states is strongly suppressed. 
As a result, the NESS is a mixture of $\ket{k_n}$ and $\ket{0}$ only, 
and the density profiles clearly have a sinusoidal shape with wave vector $k_n$.
Nevertheless, 
small differences between the OBM (solid) and the TLM (dashed lines) are visible for the resonance relative to the state $\ket{k_1}$.
As the system size is increased (middle and bottom panels), the occupation 
of the off-resonant one-photon states is not negligible, 
thus resulting in a more complicate structure of the NESS.
It is important to note (especially for large systems) that the photon density does not undergo an exponential suppression of the type $n_i\propto\rme^{-\gamma i}$ despite the photon leakage occurs extensively and the array is refilled just from one end.

\subsection{Many-photon resonances}
\label{ssec:mpr}

\begin{figure}[!t]
  \includegraphics[width=0.45\textwidth]{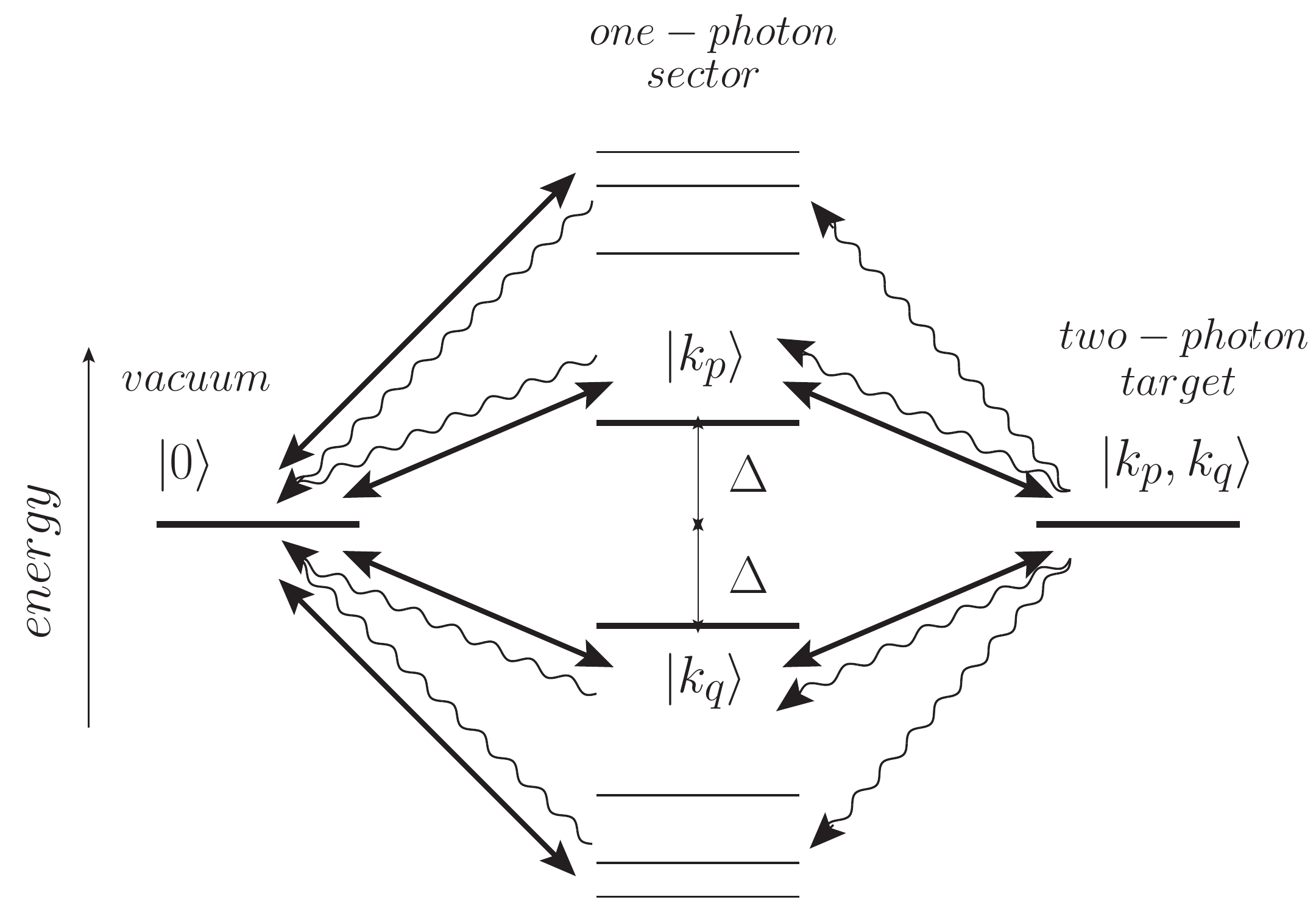}
  \caption{Typical level scheme on resonance with a two-body state $\ket{k_p,k_q}$. 
    The solid arrows stand for coherent Hamiltonian couplings, 
    while wavy arrows indicate the decay induced by the Lindbladian.}
  \label{fig:2phex2}
\end{figure}

More interesting is the characterization of the NESS in correspondence of the $N$-photon peaks. 
In this case the NESS will be a mixture of one-photon states and many-body states.
Our aim is to quantify the presence of many-body states in the NESS and to study 
their signatures on the observables.
We start analyzing the excitation of a generic two-photon state $\ket{k_p, k_q}$.
Considering Eq.~\eqref{sim}, the generic two-photon eigenstate of $\hat \h_0$ can be written as
\be
\ket{k_p,k_q}=\frac1{M+1}\sum_{i,j=1}^M f_{k_p,k_q}(i,j) \ \ada_i\ada_j\ket{0},
\ee
with $f_{k_p,k_q}(i,j) = \sgn(i - j) [ \sin(k_p i)\sin(k_q j) - \sin(k_p j)\sin(k_q i) ]$.
When the two-photon resonance condition is satisfied, 
\be
\label{restwo}
\omega_p = \frac12 [ \E(k_p)+\E(k_q) ] ,
\ee
the two-body state $\ket{k_p,k_q}$ cannot be directly excited by the driving.
The occupation of $\ket{k_p,k_q}$ is the result of the sequential absorption of two photons 
by the off-resonance one-body states $\ket{k_p}$ and $\ket{k_q}$. 
In fact the one-photon states $\ket{k_n}$ are coupled to the two-photon states $\ket{k_p,k_q}$ 
with a matrix element 
\ba
A_{k_n,k_p,k_q} & = &  \braket{k_n|\hat \h|k_p,k_q} \\
\nonumber & = & F \sqrt{\frac2{M+1}} \Big( \delta_{k_n,k_p} \sin k_q - \delta_{k_n,k_q} \sin k_p \Big) .
\ea
It is interesting to note that despite the rich structure of the two-photon state $\ket{k_p,k_q}$ in momentum space, shining only the first cavity (or equivalently the last one), implies that $\ket{k_p,k_q}$ 
is directly (coherently) coupled only to the states $\ket{k_p}$ and $\ket{k_q}$ (see App.~\ref{app:effmod}). 
On the other hand, the Lindbladian terms in the master equation~\eqref{lindblad}, 
allow (incoherent) transition from $\ket{k_p,k_q}$ to different one-body states. 
The situation is depicted in Fig.~\ref{fig:2phex2}.

\begin{figure}[htbp]
\centering
{\includegraphics[width=.4\textwidth]{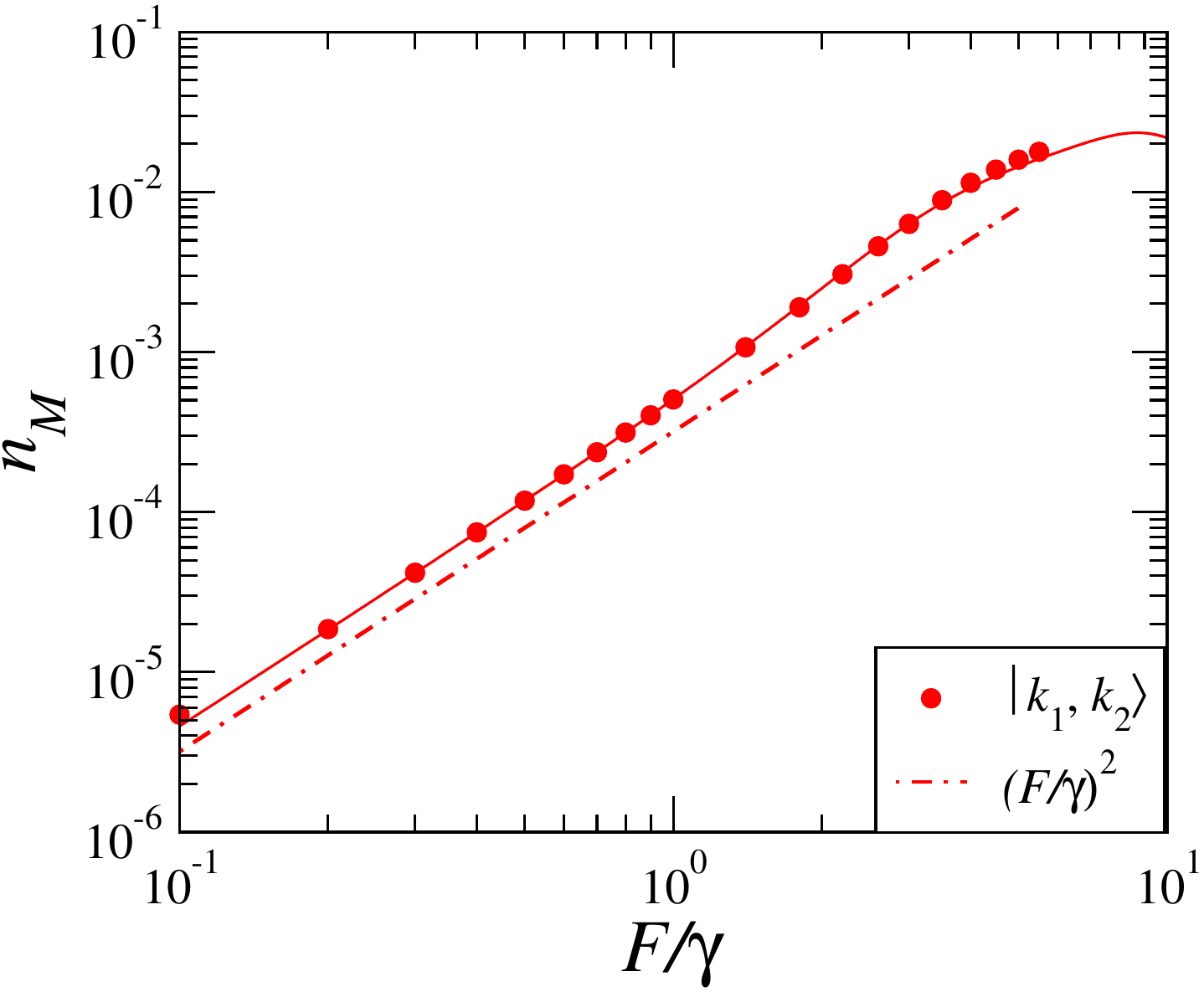}} \\ 
{\includegraphics[width=.48\textwidth]{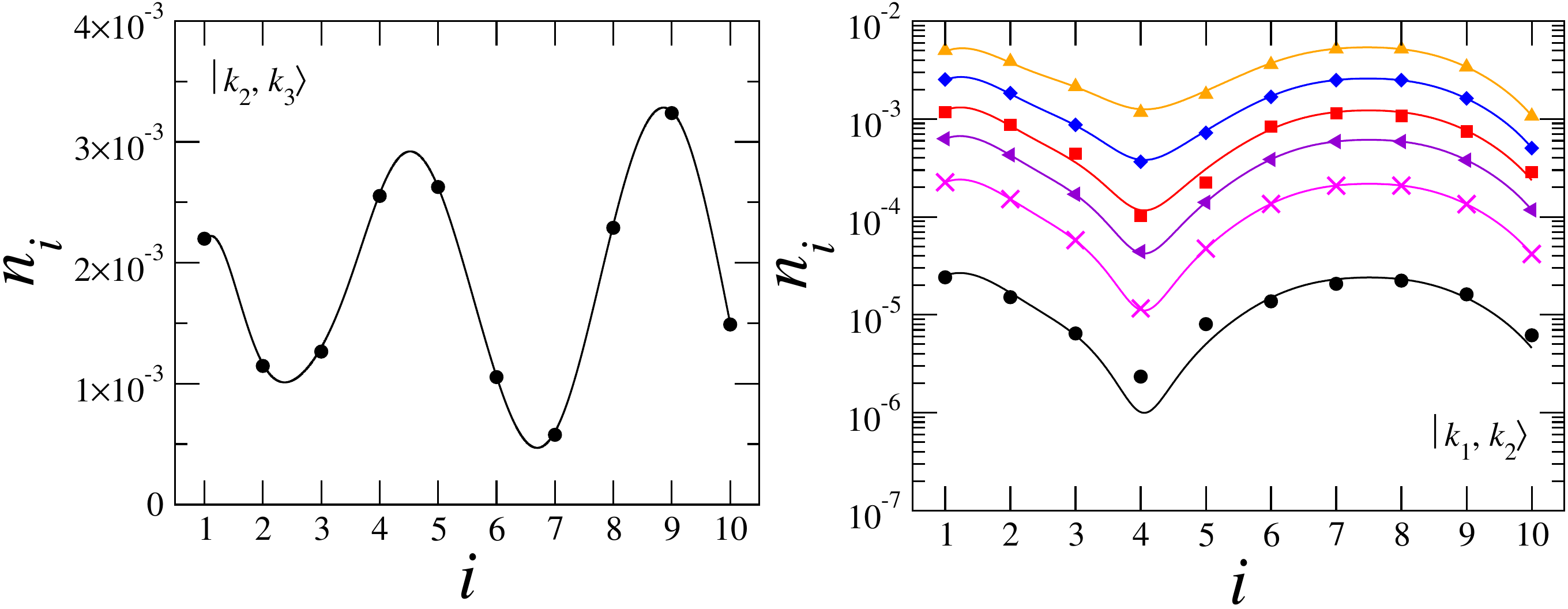}}
\caption{Top panel: the population in the $M$-th cavity as a function of the pump amplitude, 
    when the two-photon resonance condition~\eqref{restwo} is satisfied. 
    Symbols denote the numerical data, while solid lines are the TBM predictions. 
    The dashed line indicates a behavior $n_T \propto (F/\gamma)^2$, 
    and is plotted to guide the eye. 
    The parameters are the same as in Fig.~\ref{fig:spectraleft2}.
    Bottom panels: photon occupations of each site in the NESS, 
    when the two-photon resonance condition~\eqref{restwo} is satisfied. 
    Symbols are the numerical data, while solid lines are the TBM predictions. 
    Left panel: $F/\gamma=1$ and the target state is $\ket{k_2, k_3}$. 
    Right Panel: $F/\gamma=0.1, 0.3, 0.5, 0.7, 1, 1.4$ (from bottom to top respectively) 
    and the target state is $\ket{k_1, k_2}$. 
    In both panels, the remaining parameters are set as in Fig.~\ref{fig:spectraleft2}.}
\label{fig:popdens2ph}
\end{figure}

In the top panel of Fig.~\ref{fig:popdens2ph} the population in the $M$-th cavity is plotted as a function of the driving strength. 
The numerics is compared with the results of an effective model which involves all the one-body states and the target two-photon state $\ket{k_p,k_q}$.
We will refer to this model as the \emph{two-body model} (TBM).
As expected, for weak driving strength, the contribution to the population in the $M$-th cavity
is almost completely given by the sum of the populations of the single-particle states, 
and then $n_M \propto (F/\gamma)^2$. For $F/\gamma > 1$ the population in the $\ket{k_p,k_q}$ state 
becomes not negligible, thus resulting in a more complex behaviour of   which deviates from $n_M \propto (F/\gamma)^2$. 
We note that our theoretical predictions are in good agreement with the numerical data 
for all the probed values of $F/\gamma$.
However, for $F/\gamma\gg1$, any kind of effective model is expected to fail 
because of the not negligible excitation of the states not included in the model.
Also the local density in the NESS (see bottom panels of Fig.~\ref{fig:popdens2ph}) is captured by the TBM. 

In order to disclose information about the correlations in the NESS, 
we studied the normalized two-body function
\be
\label{corr2}
g^{(2)}(i,j) = \frac{\braket{\ada_i\ada_j\aaa_i\aaa_j}}{\braket{\ada_i\aaa_i}\braket{\ada_j\aaa_j}}. 
\ee
Such quantity is directly deducible from photocorrelation signals measurement. 
Furthermore, being the $g^{(2)}$ a statistical normalized quantity, 
this is of particular interest in the case of weak laser strength, when the number of photons 
in the array is very small.
Supposing that the only populated two-body state is $\ket{k_p,k_q}$, 
the normalized two-body function reads as
\be
\label{g2ness}
g^{(2)}(i,j) =\frac{\rho_{2ph}^{\rm \scriptscriptstyle NESS}}{n_i \  n_j} \left(\frac2{M+1}\right)^2 f^2_{k_p,k_q}(i,j),
\ee
where $\rho_{2ph}^{\rm \scriptscriptstyle NESS}= \braket{k_p,k_q|\rho^{\rm \scriptscriptstyle NESS}|k_p,k_q}$ 
is the population of the target two-body state in the NESS and $n_{i(j)}$ are the local densities. 
As is clear from Eq.~\eqref{g2ness}, the two-body function is directly related 
to the correlations in the target state.

\begin{figure}[!t]
  \includegraphics[width=0.48\textwidth]{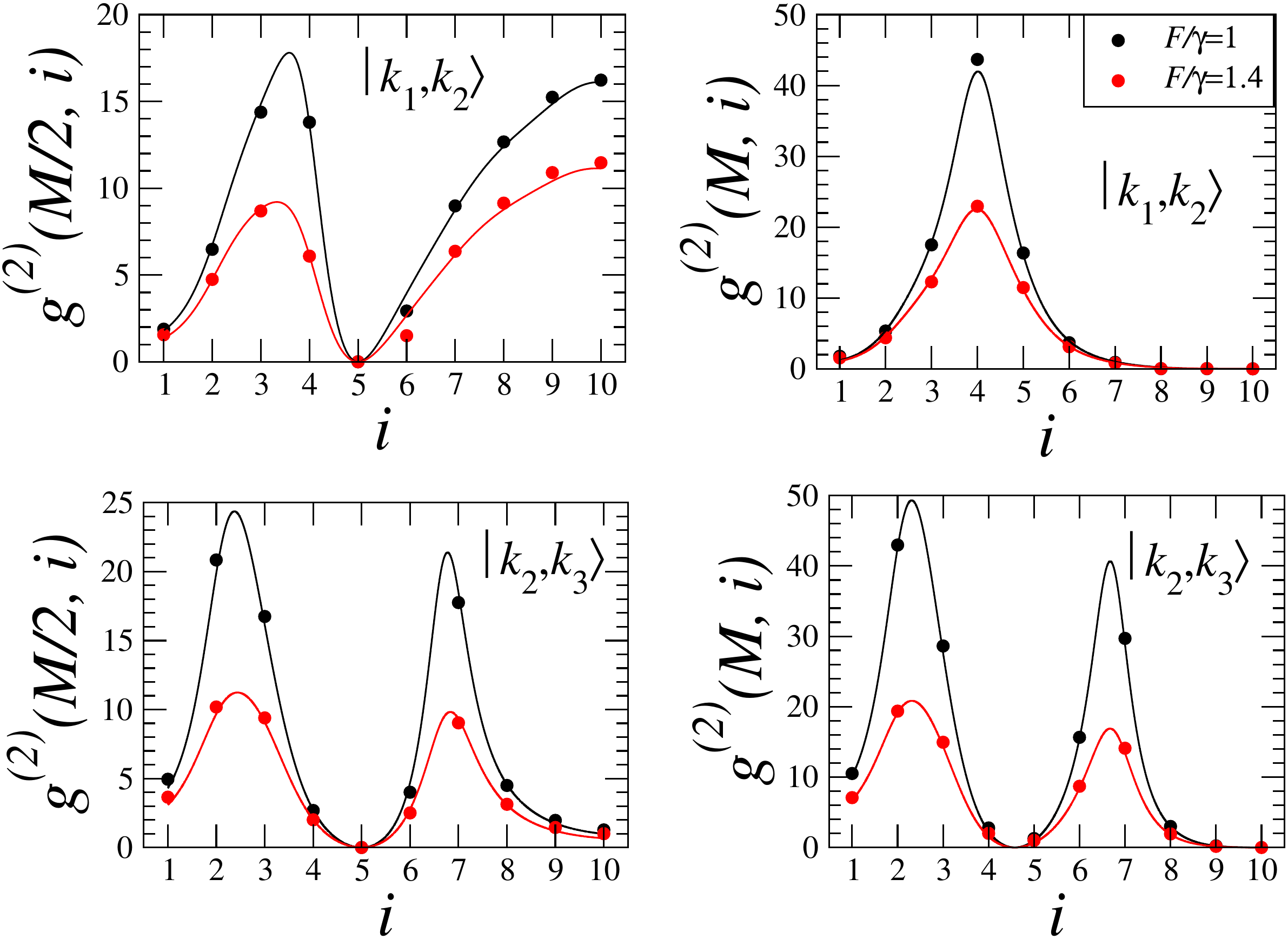}
  \caption{Normalized two-body correlation function 
    when the two-photon resonance condition~\eqref{restwo} is satisfied. 
    Symbols denote the numerical data, while solid lines are the results of the TBM. 
    The data are taken for two values of $F/\gamma$, as indicated in the legend. 
    Top panels refer to the target state $\ket{k_1, k_2}$, while the lower ones to $\ket{k_2, k_3}$. 
    The remaining parameters are set as in Fig.~\ref{fig:spectraleft2}.
    On the left plots the $g^{(2)}$ function is plotted starting from the center 
    of the array, while on the right plot we considered the $M$-th cavity as reference.}
  \label{fig:corr2ph2}
\end{figure}

The results are shown in Fig.~\ref{fig:corr2ph2}. 
for different target states and for different values of $F/\gamma$. 
Specifically, when the two-body correlation function is plotted 
taking as reference the center of the array (left panels), the auto-correlation displays 
perfect antibunching $g^{(2)}(M/2,M/2)=0$ as expected in TG limit, 
while the cross-correlations show an oscillatory behavior well captured by the TBM.
In the spirit of a transport setup, we also considered as a reference the end of the array (right panels).
Also in this case the auto-correlation of the light in the last cavity 
reveals a perfectly antibunched statistics $g^{(2)}(M,M)=0$,
and the numerical data are in good agreement with those of the effective model.

\section{Transport at finite interaction}
\label{sec:finite}
We now turn to the situation where the on-site nonlinearity is finite ($U/J<\infty$).
For this purpose it is instructive to start from the non-interacting case ($U/J=0$), which is discussed in Sec.~\ref{ssec:linear}.
This will be helpful to understand the physics at finite interactions, as analyzed in detail in Sec.~\ref{ssec:fin}, and its relation with the findings in the TG limit.

\subsection{Linear case}
\label{ssec:linear}
The case of a linear chain of cavities is integrable because both the Hamiltonian and the Lindbladian 
are quadratic in the photon creation and annihilation operators. 
The equations of motion of the needed observables are identical to the corresponding classical equations (see for example Ref.~\cite{carusotto2013}) and can be solved for the NESS 
using the fourth-order Runge-Kutta method. 
The total population in the NESS and the population in the $M$-th cavity of the chain, 
are displayed in Fig.~\ref{fig:grid} as a function of laser detuning.
In the range shown, the visible single-particle states 
(whose energy does not depend on $U/J$) are $\ket{k_5}$ and $\ket{k_6}$. 
In the free case, the states $\ket{k_n}$, $\ket{2:k_n}$, $\ket{3:k_n}$, $\ldots$ (the notation $\ket{N:k_n}$ indicates the state with $N$ particles in the $k_n$ orbital)
can be all resonantly excited (dependently on the driving strength $F/\gamma$) 
at $\omega_p=\E(k_n)$. 
This means that, on resonance, the NESS will be a mixture of one-body and (factorizable) 
many-body states composed of photons with the same momenta: 
the photon-blockade is absent because of the harmonicity of the resulting spectrum.

It is important to note that the excitation of many-body states composed by photons 
with different momenta is strongly suppressed because of a destructive interference phenomena 
in the excitation pathways. The two main processes involved in the excitation of the generic 
two-photon state $\ket{1:k_p;1:k_q}$ with $k_p\neq k_q$, 
are $\ket{0}\rightarrow\ket{k_p}\rightarrow\ket{1:k_p;1:k_q}$ and $\ket{0}\rightarrow\ket{k_q}\rightarrow\ket{1:k_p;1:k_q}$.
These can occur with the same probability, but with an amplitude carrying opposite sign. 
This is due to the fact that when the laser is resonant with the two-photon state $\ket{1:k_p;1:k_q}$ we get a level scheme similar to Fig.~\ref{fig:2phex2} where the state $\ket{k_p,k_q}$ should be replaced with the state $\ket{1:k_p;1:k_q}$ and the dissipation allows (incoherent) transition from the target two-body state $\ket{1:k_p;1:k_q}$ to $\ket{k_p}$ and $\ket{k_q}$ only. The one-body states have equal and opposite energy ($\pm\Delta$) with respect to the vacuum and are coupled to $\ket{1:k_p;1:k_q}$ with a matrix element 
\begin{equation}
  \begin{split}
    \bra{k_n} \hat \h  \big\vert_{U=0} \ket{1:k_p;1:k_q} & = \\
    = F \sqrt{\frac2{M+1}} \Big( \delta_{k_n,k_p}\sin k_q  &  + \delta_{k_n,k_q} \sin k_p\Big).
  \end{split}
\end{equation}
For this reason, despite the state $\ket{1:k_5;1:k_6}$ is an eigenstate of the free Hamiltonian, 
it is not visible in the spectrum of the total population (see Fig.~\ref{fig:grid}).

\subsection{Interacting case}
\label{ssec:fin}
\begin{figure}[!t]
  \includegraphics[width=0.45\textwidth]{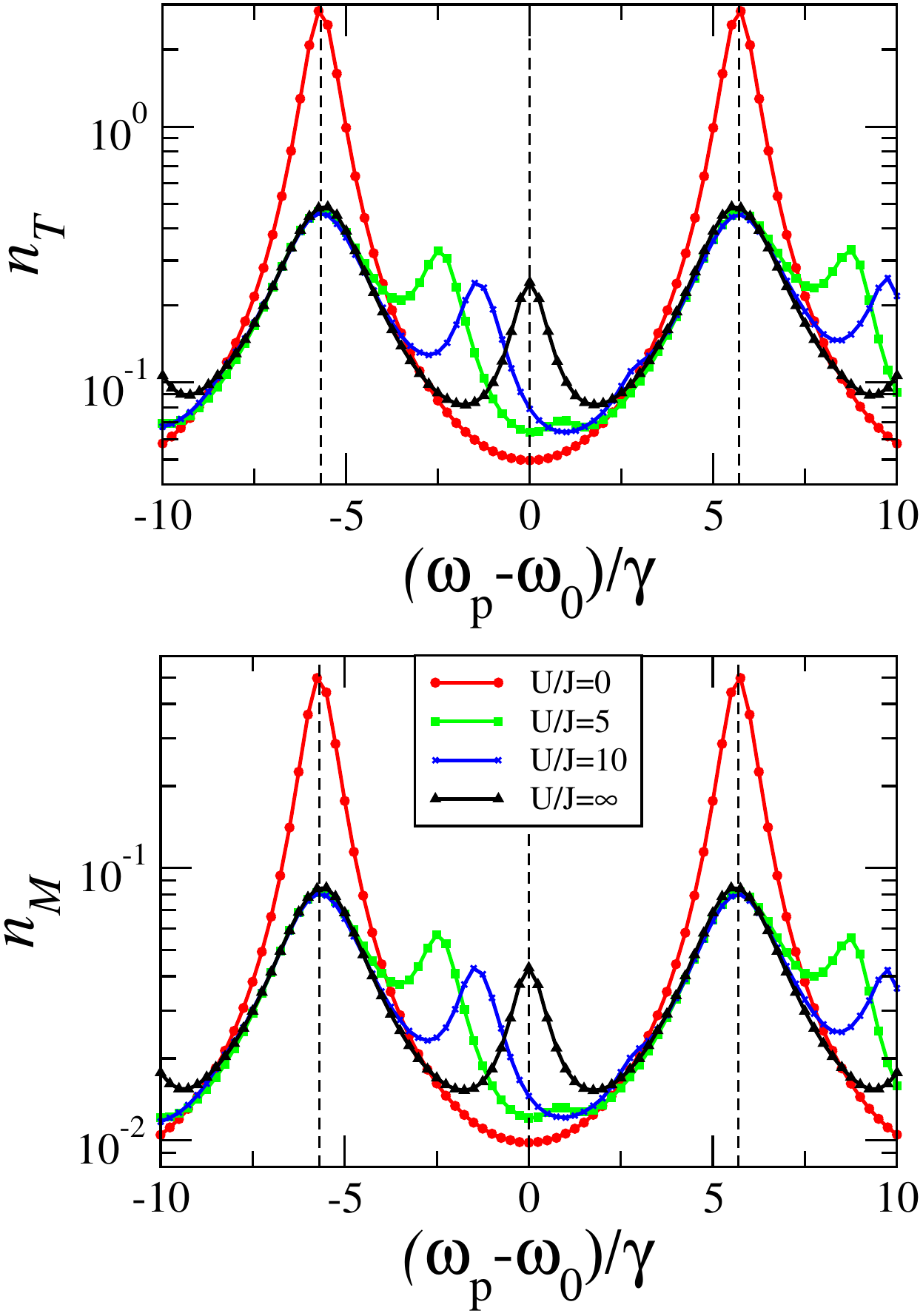}
  \caption{Total population (top panel) and population in the $M$-th cavity (bottom panel) 
    as a function of laser detuning. 
    The various symbols stand for different values of $U/J$, as indicated in the legend. 
    The dashed vertical lines are $\ket{k_5}$, $\ket{k_5, k_6}$ and $\ket{k_6}$ respectively. 
    In both panels $F/\gamma=2$, $J/\gamma=20$, $\omega_0/\gamma=1 $ and $M=10$.}
  \label{fig:grid}
\end{figure}

Switching on the nonlinearity, the resonances at $\omega_p=\E(k_n)$ are split and shifted: 
the harmonicity of the spectrum is broken and the photon-blockade takes place.
Specifically the two-photon state $\ket{2:k_n}$ is no longer an eigenstate 
of the interacting Hamiltonian. In the weakly-interacting limit ($U/J\ll1$), 
the relative resonance is blueshifted proportionally to $U/M$ 
at first order in perturbation theory~\cite{carusotto2009, grujic2012}.
Such regime is in principle difficult to study within the MPO approach, 
because of the large dimension of the local Hilbert space, but can be explored 
by means of perturbation theory in the parameter $U/J$ or using a classical discrete non-linear Schr\"odinger equation~\cite{carusotto2013}.
For larger, but still finite values of $U/J$, as shown in Fig.~\ref{fig:grid}, the resonance continues 
to be blueshifted approaching asymptotically the fermonized value~\cite{carusotto2009}.
In this regime an MPO approach gives reliable results. Interestingly we found that 
the needed local dimension for moderate driving strength ($F/\gamma=2$) 
and $M=10$ is just $d=3$ for all the values of $U/J$ that we considered (not shown).

\begin{figure}[!t]
  \includegraphics[width=0.45\textwidth]{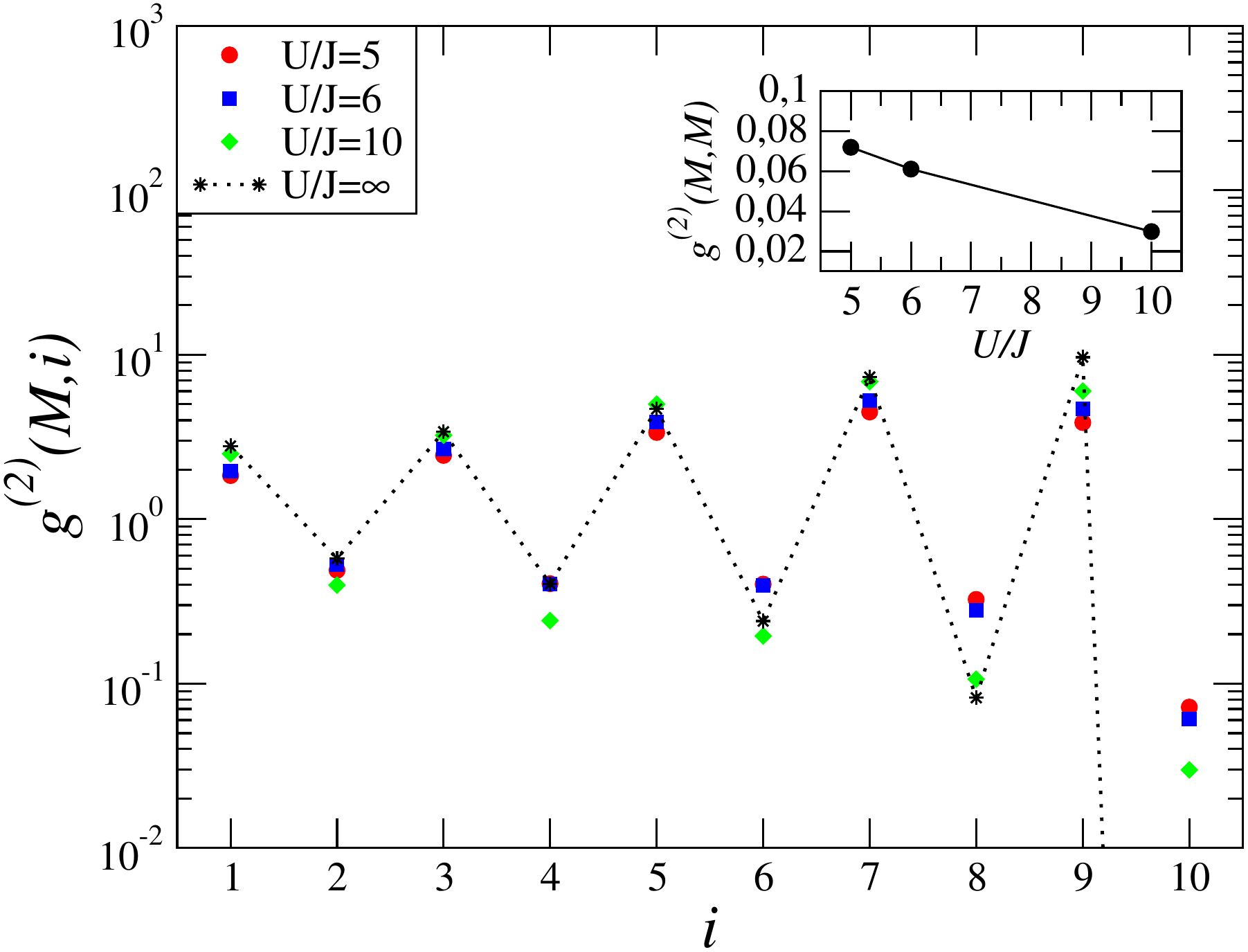}
  \caption{The normalized two-body correlation function on resonance with the $U/J$ dependent two-photon
    peak (see Fig.~\ref{fig:grid}). Here the cavity on the opposite side
    with respect to the driving laser is taken as reference.
    In the inset the auto-correlation of the $M$-th cavity is shown as a function of $U/J$. 
    The parameters are set as in Fig.~\ref{fig:grid}.}
  \label{fig:g2ufinite}
\end{figure}

In the top panel of Fig.~\ref{fig:grid} the spectra of the total number of photons in the NESS $n_T=\sum_{i=1}^M\braket{\ada_i\aaa_i}$ is shown.
We note that, being the eigenstates of $\h_0$ extended independently from $U/J$, 
it exhibits all the features of the population in the $M$-th cavity (bottom panel). 

The degree of nonlinearity of the system also influence the statistics of the output radiation.
In Fig.~\ref{fig:g2ufinite} we show the behavior of $g^{(2)}$ for different values 
of the Kerr nonlinearity $U/J$ (we keep the driving on resonance with the $U/J$ dependent 
two-photon peak of Fig.~\ref{fig:grid}).
The oscillatory behavior of the cross-correlations that we found in the TG limit 
persists also at finite interaction.
The auto-correlation of the light in the $M$-th cavity exhibits antibunching ($g^{(2)}(M,M)<1$) 
for all the value of $U/J$ probed (see inset of Fig.~\ref{fig:g2ufinite}). 
Such antibunching is more pronounced as the nonlinearity is increased.
For a detailed discussion about how the light statistics is related 
to the nonlinearity strength in CCAs see Ref.~\cite{grujic2013}.

\section{Conclusions}
\label{sec:conclusions}

In this work we studied an array of coupled nonlinear cavities subjected to dissipation and driven, at one end, by a 
coherent source. By means of numerical simulations and input-output formalism we characterized the transport properties analysing 
both the populations and the correlations of the transmitted light in the non-equilibrium steady state emerging from the interplay of 
driving and dissipation.

We found that the formation of strongly correlated (many-body) states of light in the NESS determines the transport properties 
of the system we consider. Remarkably, strong correlations play a dominant role also in very large arrays (we simulated up 
to sixty cavity). This, we believe, is a nontrivial observation since the incoherent photon leakage from each cavity  is 
expected to kill coherence between different cavities. Additionally, since we refill the array just from one end (while dissipation occurs 
extensively), it is not obvious that transport in large systems can take places at all.   
We found that the transmission of photons through the array displays single- and multi-photon peaks, 
which reveal the level structure of the array resulting from the competition between photon hopping 
and Kerr-like nonlinearity. For weak driving strength only single-photon states appear in the transmission spectra.
As the pump strength is increased, resonances related to many-body states start to appear.
Notwithstanding this fact, our work establishes that {\em photon transport is controlled by many-body resonances related to 
extended states in the cavity array.}

In the Tonks-Girardeau limit, the number of transmitted photons 
on resonance with single- and multi-photon states has been fully characterized 
and the presence of strongly-correlated states of light in the steady state 
manifests in the behaviour of the two-body correlation function 
which signals a perfect antibunched statistics of the output field.
In this regime we developed some truncated effective models (based on a careful identification of the relevant degrees of freedom) 
which allow us to confirm our numerical results and considerably increase the understanding of the physics of this complex system.
The agreement is almost perfect, both for the populations 
and for the correlation functions, confirming the effectiveness of the MPO method 
for the investigation of large (one-dimensional) open quantum many-body systems.

We went beyond the Tonks-Girardeau limit. We analyzed the case of a finite Kerr nonlinearity.
Also in this case we found that the structure of the strongly correlated states in the steady state rules the transport. 
In particular, moving from the linear regime to the Tonks-Girardeau limit, the harmonic structure of the spectrum 
is progressively lost dramatically affecting (and inhibiting) the transport properties of the array. 
This effect is the generalisation to extended systems of the well known single cavity photon blockade phenomena.
Quite interesting in this respect would be the analysis of the transport in the opposite regime in which the hopping 
is dominating over the local non-linearity. Here one  expects that the spectrum of the Hamiltonian has low-lying sound-like 
modes leading, in equilibrium, to quasi-long range order. It would be very interesting to see how photon transmission is modified 
in this "superfluid" regime. Unfortunately the case of small Kerr non-linearity is difficult to handle with the numerical 
methods used here as the dimension of the local Hilbert space grows enormously. A dissipative Luttinger liquid description or the truncated-Wigner methods~\cite{sinatra2002, carusotto2005} are probably much more appropriate in this regime. 

To conclude, our analysis can be viewed as the multi-cavity generalisation of the classic experiments on photon blockade. 
By exploring these complex architectures that show a many-body photon blockade effect, as in the present paper, an additional
ingredient to tune and control photon  transmission at single- or few-photon level can be realised.
Note that by driving all the cavities, one is limited by 
symmetry reasons to address only  few eigenstates of the system. On the contrary in our configuration, by suitably adjusting 
the laser-cavity detuning, it is possible to excite all the many-body spectrum of the system paving the way to the possibility 
to perform a complete system spectroscopy of the array studying transport. 

\acknowledgments{We would like to thank Andrea Mari for fruitful discussions. 
We acknowledge financial support by the ERC through the QGBE grant, 
by the Autonomous Province of Trento, 
partly through the project ``On silicon chip quantum optics for quantum computing and secure communications'' (``SiQuro''), 
by EU through the grants IP-project SIQS, and the STREP 
project THERMIQ, by Regione Toscana POR FSE 2007-2013, and by  the Italian MIUR  through  Projects RBFR12NLNA and 2010LLKJBX006 "Fenomeni quantistici collettivi: 
dai sistemi fortemente correlati ai simulatori quantistici".}


\appendix

\section{Effective models}
\label{app:effmod}

In this work we compared the results of the MPO simulations with the outcome 
of some truncated effective models.
All the results of this appendix refer to the TG limit ($U/J=\infty$).
The first model we consider takes into account only the one-particle sector 
of $\hat \h_0$ and the vacuum.
As in the main text, we will call it the \emph{one-body model} (OBM).
In this model, the vacuum $\ket{0}$ is coupled to the single-particle states $\ket{k_n}$ 
with a matrix element
\begin{equation}
  \label{spme}
  F_{k_n} = \braket{0|\hat \h|k_n} = F \sqrt{\frac2{M+1}} \sin k_n,
\end{equation}
where we used $\ket{k_n} = \sqrt{\frac{2}{M+1}} \sum_{i=1}^M \sin(k_n i) \, \ada_i\ket{0}$ 
with $k_n= n\pi/(M+1)$ and $n=1, \dots,M$ as imposed by the open boundary conditions.
As it is clear from Eq.~\eqref{spme}, in this driving scheme, all the single-particle 
states can be excited. 
The non-unitary part of the evolution of the master equation~\eqref{lindblad} 
is governed by the Lindbladian term~\eqref{lindbladianterm}. 
In this case is more convenient to work with the annihilation and creation operators 
of Bloch modes which are related the photons annihilation and creation operators 
as usual $\bbb_{k_n} = \sqrt{2/(M+1)} \sum_{i=1}^M \sin(k_n i)\aaa_i$.
Using the orthogonality relation $\frac{2}{M+1} \sum_{l=1}^M \sin(k_p l) \sin(k_q l)=\delta_{k_p,k_q}$, 
it is easy to show that 
\ba
\lio[\rho] &=& \frac\gamma2\sum_{i=1}^M(2\aaa_i\rho\ada_i-\ada_i\aaa_i\rho-\rho\ada_i\aaa_i) \cr
&&\cr
&=& \frac\gamma2\sum_{n=1}^M(2\bbb_{k_n}\rho\bda_{k_n}-\bda_{k_n}\bbb_{k_n}\rho-\rho\bda_{k_n}\bbb_{k_n}).
\ea
Also in this case all we need are the matrix elements 
$\braket{0|\bbb_{k_m}|k_n} = \braket{0|\bbb_{k_m}\bda_{k_n}|0}= \delta_{k_m,k_n}$. 
The result explains the why we worked in this different basis for the Lindbladian term.
The OBM is expected to work when the driving is weak ($F/\gamma\ll1$) 
and therefore only the single-particle states play a role in the dynamics. 
Nevertheless, if the pump is not resonant with a two-photon state so that the excitation 
of many-body states is strongly inhibited, the OBM is expected to work 
even if the driving is not weak (see, {\it e.g.}, Figs.~\ref{fig:pop1ph2} and~\ref{fig:dens1phlarge3}).
Of course any kind of effective model one could think will fail for large 
driving strength ($F/\gamma\gg1$) because at a certain point the excitation of the states not included in the model starts 
to be not negligible. A clear example is given in Fig.~\ref{fig:pop1ph2}.
When the laser is resonant with a one-photon state $\ket{k_n}$ and the others resonances 
are well separated in energy with respect to their width we can further simplify the model 
taking into account the target state and the vacuum only.
We will refer to this model as the \emph{two-level model} (TLM).
The TLM can be solved analytically for the steady state. What we get is
\ba
\braket{k_n|\rho^{\rm \scriptscriptstyle NESS}|k_n} & = & \frac{F_{k_n}^2}{2F_{k_n}^2 + (\gamma/2)^2} \\
\braket{k_n|\rho^{\rm \scriptscriptstyle NESS}|0} & = & \rmi\frac{2F_{k_n}}{\gamma} 
\bigg( \frac{2F_{k_n}^2}{2F_{k_n}^2+(\gamma/2)^2}-1 \bigg), \nonumber
\ea
where $\rho^{\rm \scriptscriptstyle NESS}$ is the NESS density matrix.
\begin{figure}[!t]
  \includegraphics[width=0.45\textwidth]{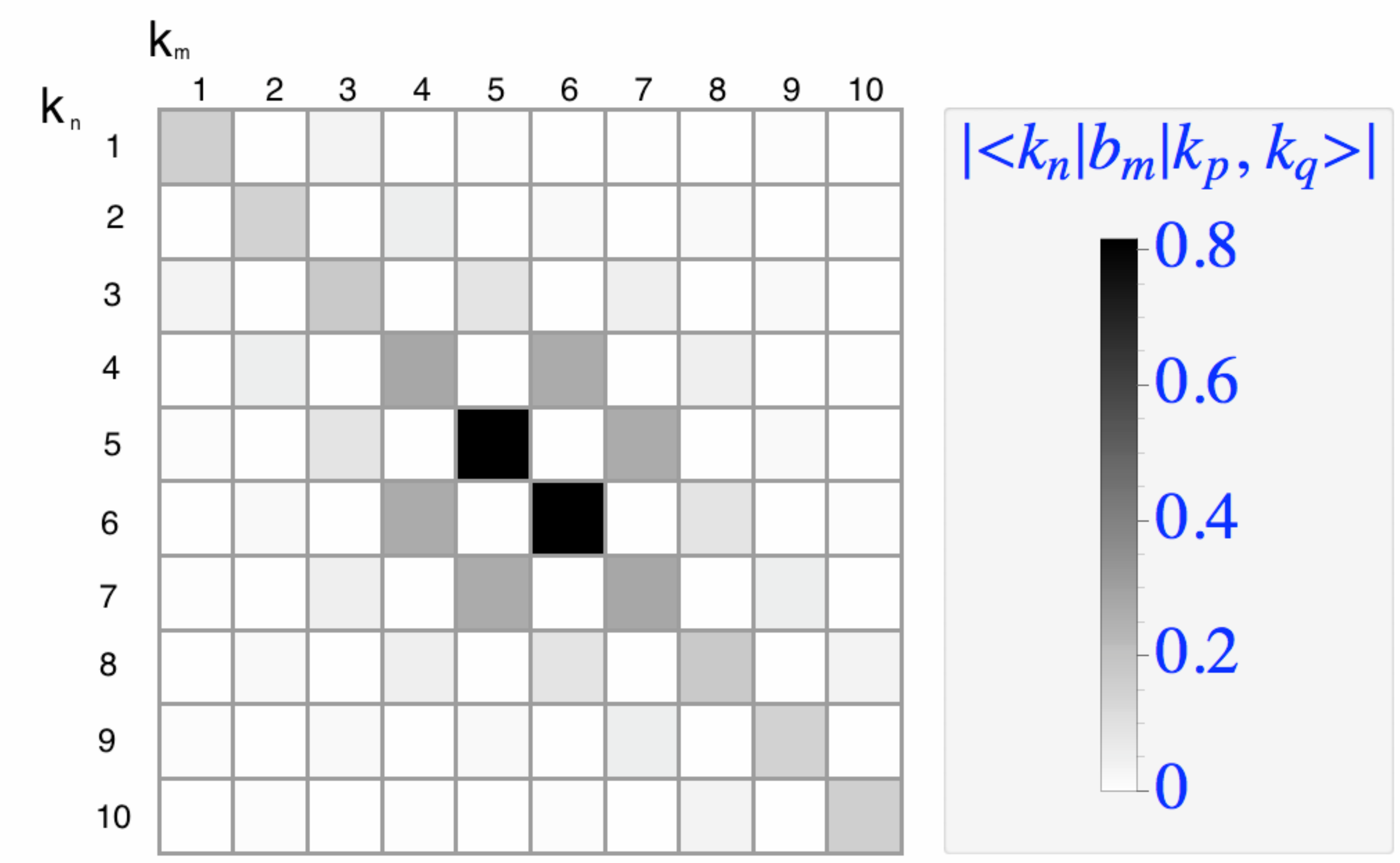}
  \caption{The absolute value of the matrix element $\braket{k_n|\bbb_m|k_p, k_q}$ 
    for the target two-photon state $\ket{k_5, k_6}$ as a function of $k_n$ and $k_m$ for $M=10$.}
  \label{fig:tomofin}
\end{figure}

When the laser is resonant with a two-photon state $\ket{k_p,k_q}$, we used 
another effective model which takes into account all the single-particle states $\ket{k_n}$, 
the target two-body state $\ket{k_p,k_q}$ and the vacuum $\ket{0}$. 
We will refer to this model as the \emph{two-body model} (TBM).
The vacuum is coupled to the single-particle states as before, 
while the two-body state is coupled to the one-body states with
\ba
A_{k_n,k_p,k_q} &=& \braket{k_n|\hat \h|k_p,k_q} \\
&=& F \sqrt{\frac2{M+1}} \left[ \delta_{k_n,k_p} \sin k_q- \delta_{k_n,k_q} \sin k_p\right]. \nonumber
\ea 
Here we explicitly used the structure of the two-body eigenstates of $\h_0$, {\it i.e.}, 
$\ket{k_p,k_q} = \frac{1}{M+1} \sum_{i,j=1}^M f_{k_p,k_q}(i,j) \ \ada_i\ada_j\ket{0}$ 
where $f_{k_p,k_q}(i,j) = \sgn(i-j) [\sin(k_p i)\sin(k_q j) - \sin(k_p j)\sin(k_q i)]$ 
and the orthogonality relation 
$\frac{2}{M+1} \sum_{l=1}^M \sin(k_p l) \sin(k_q l)=\delta_{k_p,k_q}$. 
Remarkably the state $\ket{k_p,k_q}$ is coupled \emph{only} with the states $\ket{k_p}$ and $\ket{k_q}$. 
This is unexpected because the state $\ket{k_q,k_q}$ 
has a very complicated structure in the momentum space.
Nevertheless shining only the first cavity (or equivalently the last one) 
we obtain a result very similar to the free case where 
\begin{equation}
  \begin{split}
    \bra{k_n} \hat \h \big\vert_{U=0} \ket{1:k_p;1:k_q} & = \\
    = F \sqrt{\frac2{M+1}} \left[ \delta_{k_n,k_p} \sin k_q \right. & \left. + \delta_{k_n,k_q} \sin k_p\right].
  \end{split}
\end{equation}
Analogously it can be shown that the three-photon state $\ket{k_a, k_b, k_c}$ 
is coupled only with the states $\ket{k_a, k_b}, \ket{k_b, k_c}$ and $\ket{k_a, k_c}$.

For the dissipative part of the evolution, additionally to the matrix elements 
evaluated above, we need to compute
\begin{widetext}
  \ba
  \label{mele4}
  \braket{k_n|\bbb_m|k_p, k_q} &=& \left( \frac{2}{M+1} \right)^2 \frac12 \sum_{i,j,l,s=1}^M \sin(k_nl) \sin(k_m s) f_{k_p,k_q} (i,j) \braket{0|\aaa_l\aaa_s\ada_i\ada_j|0} \cr
  &&\cr
  &=& \left( \frac{2}{M+1} \right)^2 \sum_{i,j=1}^M \sin(k_n i)\sin(k_m j)f_{k_p,k_q}(i,j). 
  \ea
\end{widetext}  
Expression~\eqref{mele4} cannot be simplified further. 
It keeps track of the very rich distribution of $\ket{k_p,k_q}$ in the momentum space. 
As it is shown in Fig.~\ref{fig:tomofin}, in contrast to the Hamiltonian evolution, 
the dissipative dynamics couples incoherently the two-photon state $\ket{k_p,k_q}$ 
not only to $\ket{k_p}$ and $\ket{k_q}$. The typical level scheme is shown in Fig.~\ref{fig:2phex2}.

\section{Remarks on the matrix-product-operator approach}
\label{app:mpo}

\begin{figure}[!t]
  \includegraphics[width=0.45\textwidth]{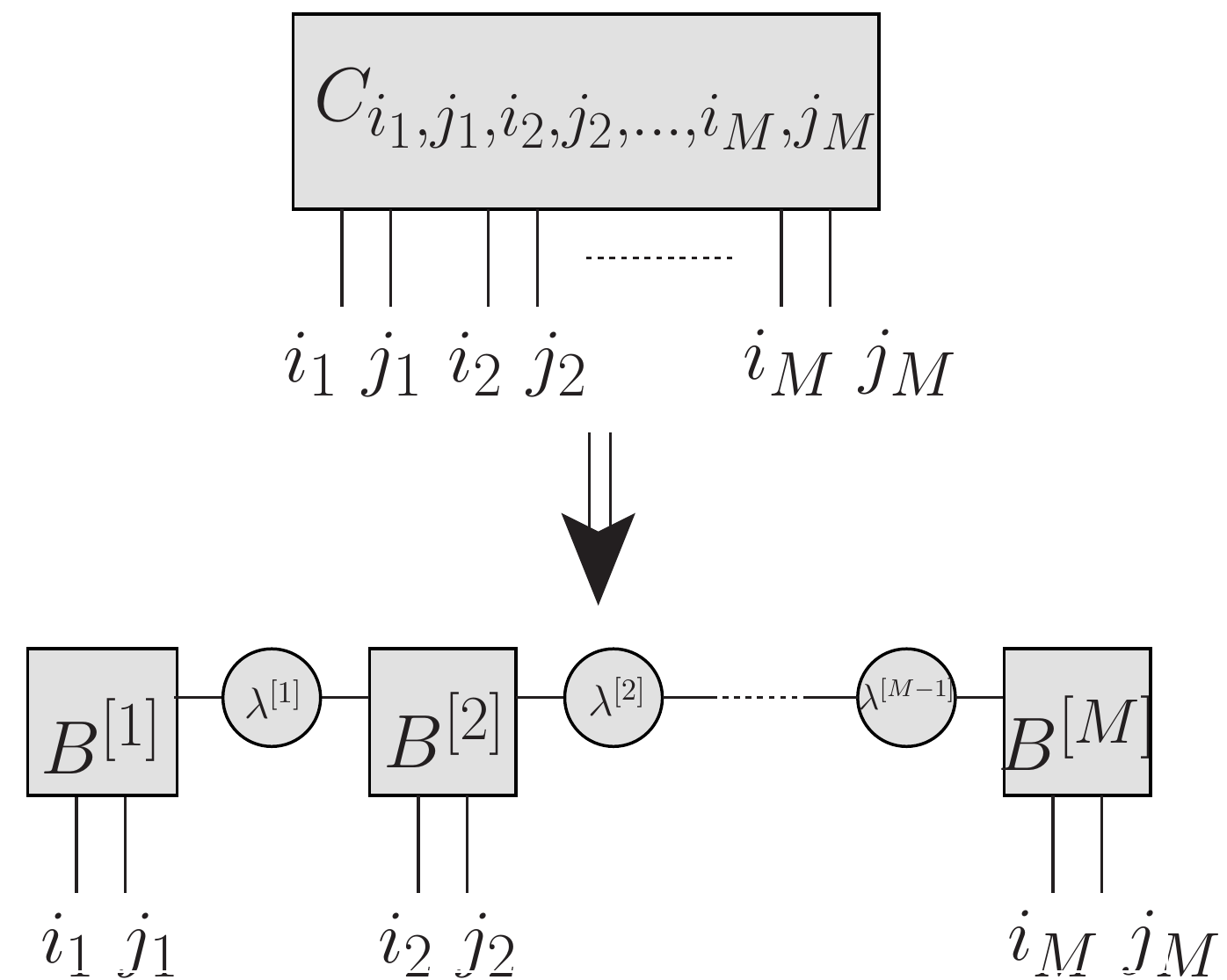}
  \caption{Graphical representation of the density matrix in a tensor-network language. 
    Each block is a multi-index tensor, open links represent the free indexes, 
    while connected links stand for the contracted indexes.}
  \label{fig:MPO}
\end{figure}

\begin{figure}[!t]
  \includegraphics[width=0.45\textwidth]{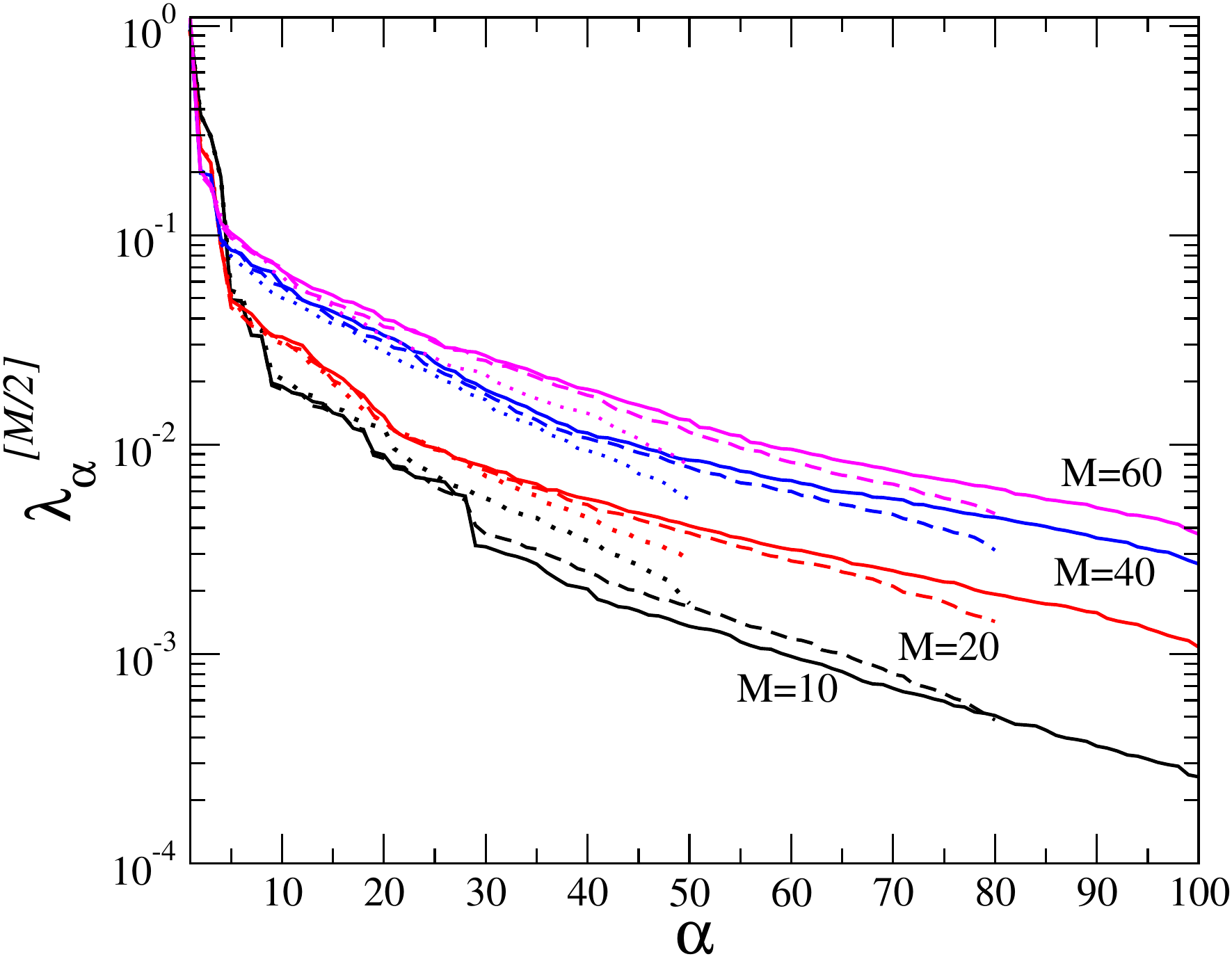}
  \caption{The spectrum of the Schmidt coefficients $\lambda_\alpha^{[M/2]}$ for a symmetric bipartition. 
    We consider various system sizes and different bond-link dimensions:
    $\chi = 50$ (dotted), $80$ (dashed), $100$ (continuous lines). 
    The laser is resonant with the single-particle state $\ket{k_{M/2}}$ and 
    the parameters are $U/J=\infty, J/\gamma=20, \omega_0/\gamma=1$ and $F/\gamma=1$.}
  \label{fig:schspec}
\end{figure}

\begin{figure}[!t]
  \includegraphics[width=0.5\textwidth]{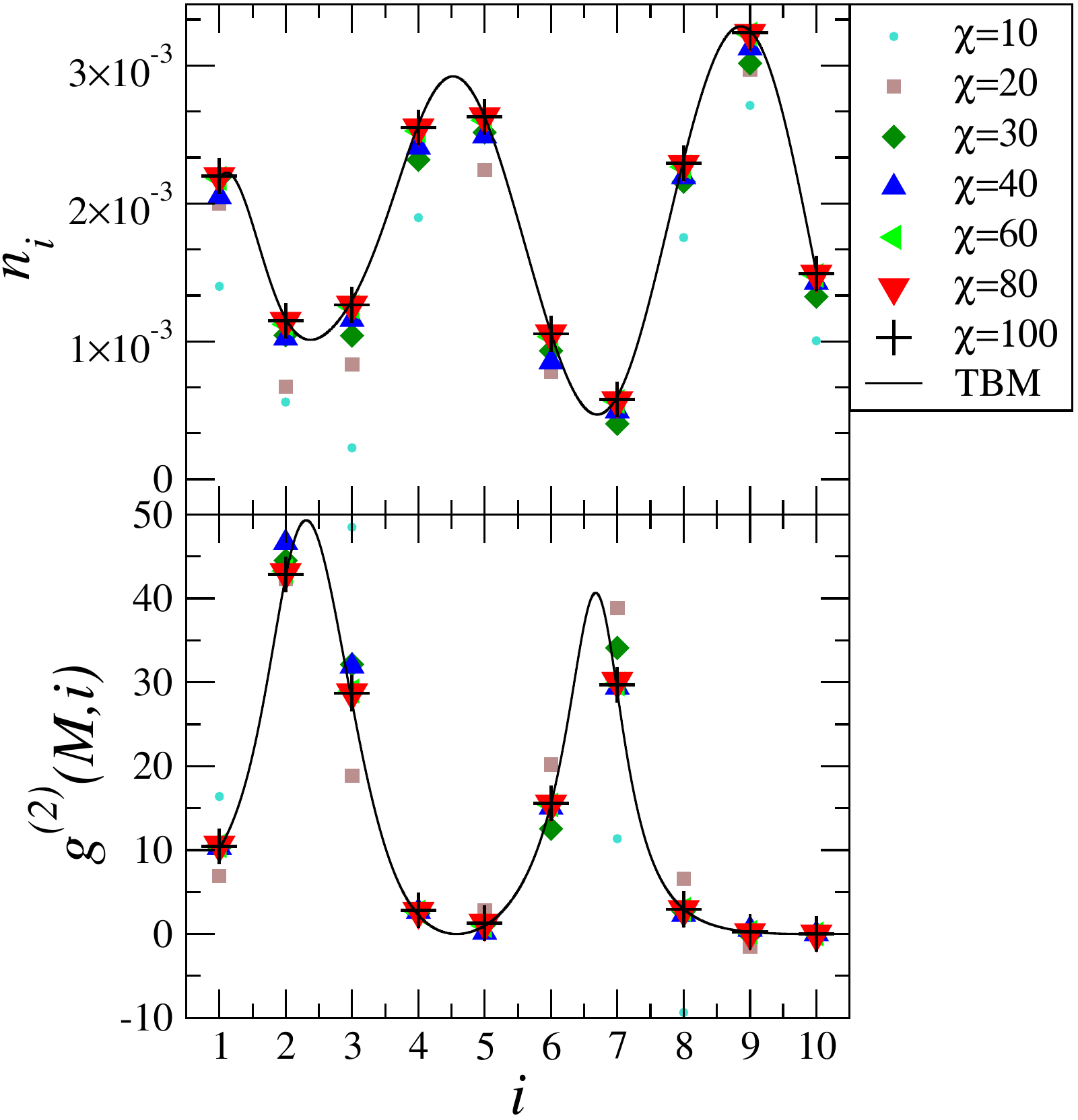}
  \caption{Local density (top panel) and normalized two-body function (lower panel) for different values of the bond link dimension $\chi$. As indicated in the legend the solid lines are the predictions of the TBM (see App.~\ref{app:effmod}).  
  Here the laser is resonant with the two-photon state $\ket{k_2, k_3}$ of an array of $M=10$ cavities. The parameters are $U/J=\infty, J/\gamma=20, \omega_0/\gamma=1$ and $F/\gamma=1$. 
The data for $\chi=100$ are shown in the left lower panel of Fig.~\ref{fig:popdens2ph} (local density) and in the right lower panel of Fig.~\ref{fig:corr2ph2} (two-body function).}
  \label{fig:bondlink_conv}
\end{figure}

\begin{figure}[!t]
  \includegraphics[width=0.45\textwidth]{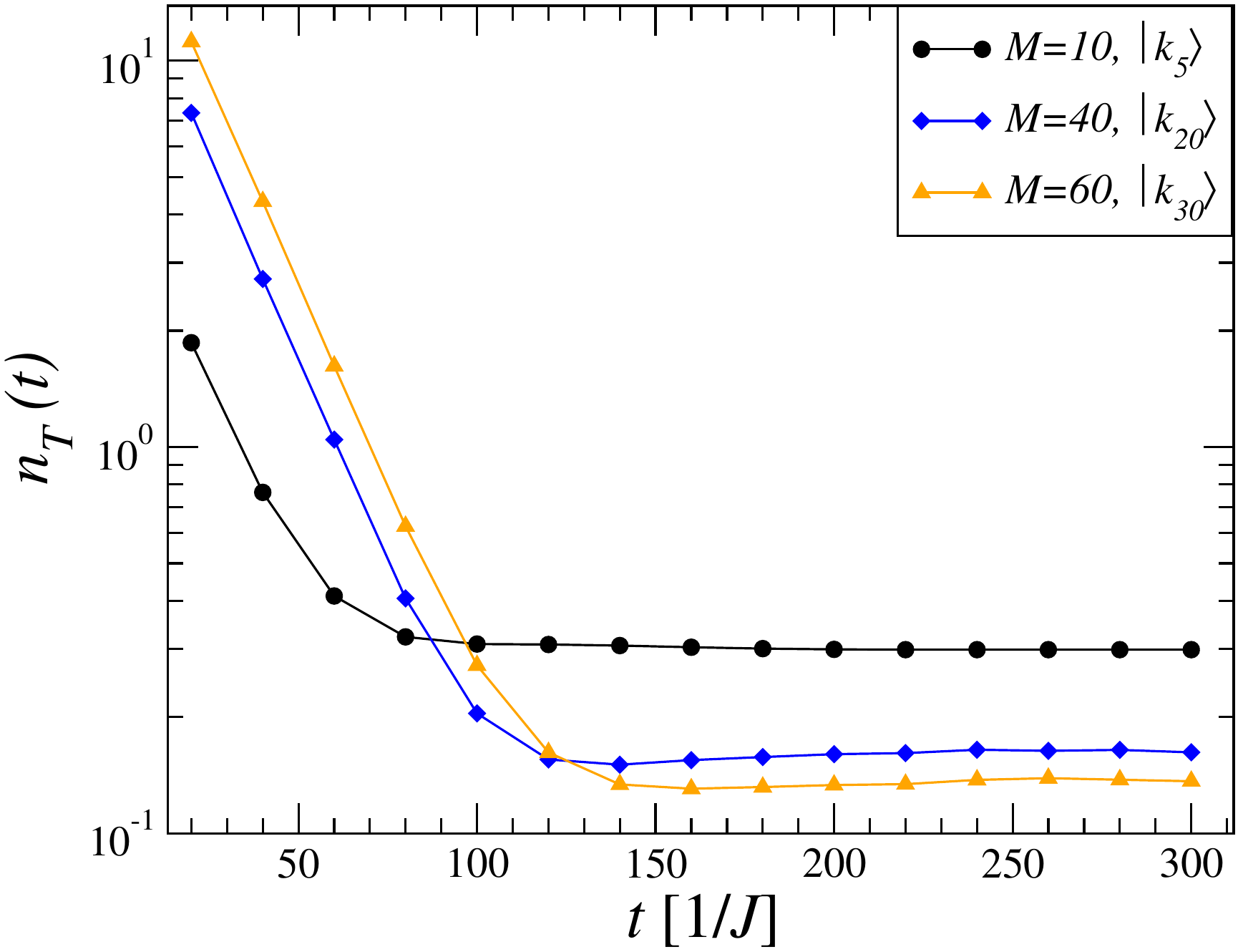}
  \caption{Evolution toward the NESS value of the total population for different system sizes 
    and target states, as indicated in the legend. 
    The initial state is random and the parameters are set as in Fig.~\ref{fig:schspec}.}
  \label{fig:convergenceness}
\end{figure}

This summary about the superoperator renormalization group technique 
is based on the appendix given in Ref.~\cite{lauchli}.
For the time-evolution toward the NESS of the density matrix we exploit an algorithm 
based on the time-evolving block decimation (TEBD) scheme~\cite{vidal2003,vidal2004} 
extended to the open systems~\cite{zwolak2004,verstraete2004}. 
In this framework the density matrix for an array of $M$ cavities and open boundary conditions 
\be
\rho = \sum_{i_\alpha, j_\alpha=1}^d \ C_{i_1 \cdots i_L, \, j_1 \cdots j_M} \, || i_1 \cdots i_M , \, j_1 \cdots j_M \rangle \rangle
\ee
is written as a MPS in the enlarged Hilbert space of dimension $d^2$, 
where $d$ is the dimension of the local Hilbert space $\mathbb{H}$.
In the specific, a repeated application of singular value decompositions 
of the tensor $C_{i_1 \cdots i_L, \, j_1 \cdots j_M}$ 
leads to the following representation (see Fig.~\ref{fig:MPO}):
\ba
\rho &=& \sum_{i_\alpha, j_\alpha=1}^d \sum_{\alpha,\beta,\dots,\gamma=1}^{\chi} \ B_{1,\alpha}^{[1]i_1,j_1}\lambda_{\alpha}^{[1]} B_{\alpha,\beta}^{[2]i_2,j_2}\lambda_{\beta}^{[2]}\dots \cr
&&\cr
&&\dots \lambda_{\gamma}^{[M-1]}B_{\gamma,1}^{[M]i_M,j_M}  || i_1 \cdots i_M , \, j_1 \cdots j_M \rangle \rangle .
\ea
Here $|| i_1 \cdots i_M , \, j_1 \cdots j_M \rangle \rangle = \bigotimes_{a=1}^{M}\ket{i_a}\bra{j_a}$ 
represents a basis for the density matrix in the product Hilbert space 
$\mathbb{H}^{\otimes M}\otimes\mathbb{H}^{\otimes M}$. 

As explained in Ref.~\cite{zwolak2004}, if the Schmidt spectrum $\lambda_{\alpha}^{[i]}$ decays 
fast enough, it can be truncated keeping only the $\chi$ largest Schmidt values.
In our simulations we fix $\chi=100$. This choice is well justified, on the basis of 
the behavior of the Schmidt spectrum for a typical choice of parameters (see Fig.~\ref{fig:schspec}).
This is reflected on how the observables studied in this work (local density and two-body function) converge as $\chi$ is increased for a typical choice of parameters (see Fig.~\ref{fig:bondlink_conv}).
As an extension of the entanglement entropy, the operator space 
entanglement-entropy~\cite{prosen_osee, znidaric_osee} of a bipartition $A$ 
of size $l$ is straightforwardly related to the behavior of the Schmidt spectrum:
\be
{\cal S}_l = -2 \sum_{\alpha} (\lambda_\alpha^{[l]})^2 \log_2 \lambda_\alpha^{[l]}.
\ee

The time-evolution is then performed using a Suzuki-Trotter decomposition 
(at fourth-order in our simulations)
of time evolution superoperator. Once the NESS is reached, the expectation values 
of some operators are obtained in the standard way $\braket{\hat O} = {\cal Z}^{-1} \ \tr[\rho^{\rm \scriptscriptstyle NESS} \, \hat O]$ 
where ${\cal Z} = \tr \, \rho^{\rm \scriptscriptstyle NESS}$ is the partition function.
The typical evolution toward the NESS value for the observable 
$n_T = \sum_{i=1}^M\braket{\ada_i\aaa_i}$ is shown in Fig.~\ref{fig:convergenceness} 
for different system sizes.

\end{document}